\documentclass[a4paper,12pt]{article}

\pdfoutput=1 

\usepackage{jheppub} 

\usepackage{enumerate}
\usepackage{ytableau}
\usepackage{mathrsfs}
\usepackage{amsmath,amsfonts,amssymb,amsthm,graphics,graphicx,epsfig,times,bbm,verbatim}

\usepackage{geometry}
\rightline{USTC-ICTS-14-21}

\title{\boldmath On the Elliptic Genus of Three E-strings and Heterotic Strings}
\author[\S]{Wenhe Cai,}
\author[\S]{Min-xin Huang,}
\author[\dagger]{Kaiwen Sun}
\affiliation[\S]{Interdisciplinary Center for Theoretical Study, Department of Modern Physics, University of Science and Technology of China, 96 Jinzhai Road, Hefei, Anhui 230026, China}
\affiliation[\dagger]{Department of Mathematics, University of Science and Technology of China, 96 Jinzhai Road,  Hefei, Anhui 230026, China}

\emailAdd{edlov@mail.ustc.edu.cn}
\emailAdd{minxin@ustc.edu.cn}
\emailAdd{skw@mail.ustc.edu.cn}

\begin{document}

\abstract{A precise formula for the elliptic genus of three E-strings is presented. The related refined free energy coincides with the result calculated from topological string on local half K3 Calabi-Yau threefold up to genus twelve. The elliptic genus of three heterotic strings computed from M9 domain walls matches with the result from orbifold formula to high orders. This confirms the $n=3$ case of the recent conjecture that $n$ pairs of E-strings can recombine into $n$ heterotic strings.}

\maketitle
\flushbottom
\bibliographystyle{unsrt}
\ytableausetup{mathmode, boxsize=0.4em}
\section{Introduction and summary}
E-string theory is a superconformal field theory in six dimension with $(1,0)$ supersymmetry \cite{Witten:1995gx}\cite{Ganor:1996mu}\cite{Seiberg:1996vs}. In the Ho\v rava-Witten picture of $E_8\times E_8$ heterotic string theory, E-strings can be realized by M2-branes stretched between a M5-brane and a M9-brane. The toroidally compactification of this theory is related to many interesting supersymmetric gauge theories in four and five dimension \cite{Ganor:1996pc}. The elliptic genus of E-strings is also equivalent to the topological string partition function for local half K3 Calabi-Yau threefold \cite{Klemm:1996hh}\cite{Minahan:1998vr}. These connections have drawn continuous attention to E-strings, see e.g. \cite{Iqbal:2002}. Furthermore, the techniques for E-strings can be also applied to topological strings on compact Calabi-Yau spaces \cite{Alim:2012, Klemm:2012,Alim:2013} which are elliptic fibrations over del Pezzo surfaces.
\\
\indent One of the main goal of studying E-strings is to compute the elliptic genus, or BPS index in other word. The technique of topological string allows one to compute some low genus refined free energy for half K3 and therefore obtain the low order information of the elliptic genus of E-strings \cite{Huang:2013yta}. These refined BPS invariants can be defined rigorously in mathematics as stable pair invariants \cite{CKK} and in some cases can be shown to be equivalent to the ones from topological strings \cite{Katz:2014}. See also the work \cite{Sakai:2011xg}, which used the Seiberg-Witten curve for E-strings \cite{Mohri:2001, Eguchi:2002, Eguchi:2002nx} in the unrefined case.
\\
\indent  Recently, a new approach based on the computation of M9 domain walls are developed, which in principle may enable us to obtain the explicit expressions for the elliptic genus of $n$ E-strings \cite{Haghighat:2014pva}. This method is based on the previous work on M-strings, especially the formulas of M5 domain walls \cite{Haghighat:2013gba, Haghighat:2013tka, Hohenegger:2013}  and the knowledge of the structure of Jacobi forms \cite{Zagierbook}\cite{Dabholkar:2012nd}. In \cite{Haghighat:2014pva}, the M9 domain walls for level two partition ware determined, and an explicit all-genus formula for the elliptic genus of two E-strings was proposed and passed highly nontrivial check. It was found that unlike heterotic strings but like M-strings, E-strings form rather nontrivial bound states. The M9 domain wall blocks can also be used to construct new formulas for the elliptic genera of $n$ heterotic strings, which have completely different appearance from the known orbifold formulas. In the work, it is conjectured that $n$ pairs of E-strings can recombine to give $n$ heterotic strings (H), or $nE+nE\rightarrow nH$ for short. The cases of $n=1,2$ was checked in \cite{Haghighat:2014pva}. For the two-string case, $2E+2E\rightarrow 2H$ involves highly non-trivial identities among Jacobi forms, which has not been exactly proved yet but passed the low order checks. We intend to follow this approach to study the elliptic genus of more than two E-strings. See also \cite{Sakai:2014hsa, Ishii:2013} for some new progress on the elliptic genus of $n$ E-strings under $D_4\oplus D_4$ twist and relation to Nekrasov-type formula.\\
\indent The main focus of this paper is on three E-strings. Based on the approach in \cite{Haghighat:2014pva}, we determine the M9 domain walls for level three partition and obtain an explicit formula for the elliptic genus of three E-strings. We also calculate the refined free energy for three E-strings up to genus 12, which all coincide with the datum from topological strings on half K3 Calabi-Yau space using the modular anomaly equation. The advantage of our present formula is that it is supposed to provide an all-genus amplitude, which is impossible to achieve with the technique of modular anomaly equation so far. We also use the M9 domain walls to recover the elliptic genus of three heterotic strings. Our formula coincides with the known results of three heterotic strings up to high orders.\\
\indent Now we briefly review the brane configurations relevant to our present paper, see \cite{Haghighat:2014pva} for more details. Our setup is based on the Ho\v rava-Witten picture \cite{Horava:1996ma}. Considering M-theory compactified on $T^2\times \mathbb{R}^8\times S^1/\mathbb{Z}_2$, in which M2-branes and M5-branes wrap on $T^2$, we parametrize the torus by $X^0,X^1$, the orbifold $S^1/\mathbb{Z}$ by $X^6$ and $\mathbb{R}^8$ as $\mathbb{R}^4_{2345}\times\mathbb{R}^4_{78910} $. When the size of $S^1/\mathbb{Z}$ goes to zero, the M2-branes stretched between two M9 branes located at the fixed points of orbifold action $X^6=0,\pi$ give rise to heterotic strings. The world sheet theory of such strings carries supersymmetry $(8,0)$. In \cite{Haghighat:2013gba}, a twisted background was introduced to break down part of the supersymmetry, similar to the $\Omega$-background introduced by Nekrasov \cite{Nekrasov:2002qd} for Seiberg-Witten theory \cite{SW}. As we go around the cycles of $T^2=S^1\times S^1$, the $\mathbb{R}^4_{2345}\times\mathbb{R}^4_{78910} $ is twisted by the action of the Cartan subalgebra of the $SO(8)$ R-symmetry parametrized by $U(1)_{\epsilon_1}\times U(1)_{\epsilon_2}\times U(1)_{\epsilon_3} \times U(1)_{\epsilon_4}$:
\begin{eqnarray}
	\prod_{i=1}^4 U(1)_{\epsilon_i} & : & (z_1,z_2) \mapsto (e^{i\epsilon_1} z_1,e^{i \epsilon_2}z_2), \\
	~ & : & (w_1,w_2) \mapsto (e^{i \epsilon_3}w_1,e^{i \epsilon_4}w_2),
\end{eqnarray}
where we impose the following constraint
\begin{equation}
	\epsilon_1 + \epsilon_2 + \epsilon_3 + \epsilon_4 = 0,
\end{equation}
such that the remaining supersymmetry is $(2,0)$. The elliptic genus of $n$ heterotic strings wrapping the $T^2$ is given by
\begin{equation}\label{eq:ZnHstr}
	Z^{\textrm{Het}}_n(\tau, \epsilon_1, \epsilon_2, \epsilon_3, \epsilon_4, \vec{m}_{E_8\times E_8})=\textrm{Tr}_{\textrm{R}} (-1)^F q^{H_L} \bar{q}^{H_R} \prod_a x_a^{K_a},
\end{equation}
where $\tau$ denotes the complex structure of the torus and $K_a$ denote the Cartan generators associated with general supersymmetry preserving $SO_R(8)$ spacetime twists and $E_8 \times E_8$ fugacities.\\
\indent On the other hand, E-strings are realized by M2-branes suspended between M5-brane and M9-brane. In the above setup and under the same twist, the elliptic genus of E-strings can be written as
\begin{equation}\label{eq:ZnEstr}
	Z_n^{\textrm{E-str}}(\tau, \epsilon_1,\epsilon_2,\vec{m}_{E_8}).
\end{equation}
The reason why this elliptic genus does not depend on $\epsilon_3,\epsilon_4$ is that E-string theory only enjoys a $SU(2)$ R-symmetry which can be identified with $SU(2)_L$ in the decomposition
\begin{equation}
	Spin(4)_{78910} = SU(2)_L \times SU(2)_R,
\end{equation}
while the $U(1)$ symmetry associated to $\epsilon_3-\epsilon_4$ lies in $SU(2)_R$.\\
\indent The main goal of the present paper is to give an exact expression of Eq \eqref{eq:ZnEstr} and a new formula of Eq \eqref{eq:ZnHstr} for the $n=3$ case. The organization of this paper is as follows: in Section two, we review the main known results on elliptic genus of E-strings and heterotic strings, especially the new approach proposed in \cite{Haghighat:2014pva} which we will follow in this paper; in Section three, we determine the relevant M9 domain walls and use them to compute the elliptic genus of three E-strings and heterotic strings.\\
\indent Our convention and notation follow the paper \cite{Haghighat:2014pva}. See the Appendix of \cite{Haghighat:2014pva} for some basic knowledge of modular forms and Jacobi forms. See also the Appendix C of \cite{Huang:2013yta} for the explicit expression of the nine Weyl invariant $E_8$ Jacobi forms.
\section{Known results for E-strings and H-strings}
In this section we review some known results on E-strings and H-strings. In Section 2.1, we state the well known formula for the elliptic genus of $n$ H-strings and the relation between elliptic genus and free energy. See original paper \cite{Dijkgraaf:1996xw} for some details. In Section 2.2, we briefly summarize the approach to calculate refined free energy of E-strings from topological strings on half K3 surface. See \cite{Sakai:2011xg}\cite{Huang:2013yta} for details. In Section 2.3, we summarize the new approach proposed in \cite{Haghighat:2014pva} and the known results for the case of two strings.
\subsection{Orbifold formula for $n$ H-strings}
Unlike E-strings and M-strings, $n$ H-strings do not form bound states. The free energy of $n$ heterotic strings wrapping on $T^2$ is just the $n$-times wound single heterotic string. This allows us to compute $Z_n^{\mathrm{Het}}$ purely from $Z_1^{\mathrm{Het}}$. The elliptic genus of single heterotic string can be easily obtained by analyzing the degree of freedom on the worldsheet, we only state the result here,
\begin{equation} \label{eq:1hetstring}
	Z^{\textrm{Het}}_1 = -\frac{\Theta_{E_8}(\vec{m}_{E_8,L})\Theta_{E_8}(\vec{m}_{E_8,R})}{\eta^{12} \theta_1(\epsilon_1) \theta_1(\epsilon_2)\theta_1(\epsilon_3)\theta_1(\epsilon_4)}.
\end{equation}
See \cite{Haghighat:2014pva} for details. The full partition of heterotic strings can be expanded with respect to $Q$,
\begin{equation} \label{Hetpf}
	Z^{\textrm{Het}} = \sum_{n \geq 0} Q^n Z^{\textrm{Het}}_n,
\end{equation}
where $Q = e^{2\pi i \rho}$ with $\rho$ being the complexified K\"ahler parameter of the $T^2$ and $Z^{\textrm{Het}}_0$ is taken to be $1$. The elliptic genus and free energy are related by
\begin{equation}
	\mathcal{F}^{\textrm{Het}} = \log(Z^{\textrm{Het}}) = \sum_{n \geq 1} Q^n \mathcal{F}^{\textrm{Het}}_n.
\end{equation}
In \cite{Dijkgraaf:1996xw}, it is shown $\mathcal{F}^{\textrm{Het}}_n$ can be expressed in an elegant way in terms of $\mathcal{F}^{\textrm{Het}}_1$:
\begin{equation}
	\mathcal{F}^{\textrm{Het}}_n = T_n \mathcal{F}^{\textrm{Het}}_1,
\end{equation}
where the Hecke operator $T_n$ acts on a weak Jacobi form $f(\tau,z)$ of weight $k$ as
\begin{equation}
	T_n f(\tau,z) = n^{k-1} \sum_{\stackrel{ad=n}{a,d>0}} \frac{1}{d^k} \sum_{b~ (\textrm{mod} ~d)} f\left(\frac{a\tau+b}{d},a z\right).
\end{equation}
Therefore, the elliptic genus of E-strings is
\begin{equation} \label{eq:Hetpfres}
	Z^{\textrm{Het}} = \exp\Bigg[\sum_{n\geq 0} Q^n \frac{1}{n} \sum_{\stackrel{ad=n}{a,d>0}}\sum_{b(\textrm{mod}~d)} Z_1^{\textrm{Het}}\left(\frac{a \tau+b}{d},a \epsilon_i, a \vec{m}\right) \Bigg],
\end{equation}
Together with Eq (\ref{Hetpf}), we can compute all $Z_n^{\mathrm{Het}}$. Hecke operator $T_n$ transforms an index $m$ Jacobi form to an index $nm$ Jacobi form and keep the modular weight unchanged. This guarantees the essential properties for the elliptic genus of $n$ strings. We will use this formula to calculate $Z_3^{\mathrm{Het}}$ and match it with our results from domain wall method.
\subsection{Results from topological strings on half K3 surface}
The elliptic genus of E-strings is equivalent to the topological strings partition function for half K3 surface. Half K3 surface is a non-compact Calabi-Yau threefold in which the half K3 surface appears as a divisor. It can be embedded in an elliptic fibration over Hirzebruch surface. This surface can also be described as the del Pezzo surface $\mathcal{B}_9$ obtained by blowing up $\mathbb{P}^2$ at 9 points. In \cite{Huang:2013yta}, a refinement of HST modular anomaly equation \cite{Hosono:1999qc} was proposed. This refined modular anomaly equation make it possible to compute the refined GV invariants for rather general homology classes $n_bp+df$ in the half K3 surface, in which the wrapping number $n_b$ and $d$ on the base $p$ and fiber $f$ correspond to the winding and momentum number of the E-strings. By the technique of topological strings, the refined free energy of E-strings can be computed to very high genus (but not all genus). Since we only use the datum of refined free energy in this work, we just briefly review their method in the following. See \cite{Sakai:2011xg}\cite{Huang:2013yta} for details.\\
\indent The refined topological string partition function and E-string elliptic genus are related by
\begin{equation}
Z(\epsilon_1,\epsilon_2) = \sum_{n=0}^\infty Q_{t}^{\,n} Z^{E-str}_{n}(\tau;\epsilon_1,\epsilon_2,\vec{m}_{E_8}),
\end{equation}
where $Q=e^{2\pi i t}$ with $t$ being the string tension. The coefficient of $Q^n$ counts the states coming from $n$ E-strings wrapping the torus. By taking the logarithm of the partition function, we can obtain the total free energy:
\begin{equation}
\mathcal{F} =  \log\left(Z(\epsilon_1,\epsilon_2)\right).
\end{equation}
In the refined case, the free energy can be expanded in the following form:
\begin{equation}
\mathcal{F} = \sum_{n \geq 0}\sum_{g\geq 0}\sum_{\ell \geq 0} Q^{n} (-\epsilon_1\epsilon_2)^{g-1}(\epsilon_1+\epsilon_2)^{2\ell}\mathcal{F}_{n,g,\ell}.
\end{equation}
The refined free energy $\mathcal{F}_{n,g,\ell}$ satisfy the following refined modular anomaly equation,
\begin{eqnarray}
\partial_{E_2}\mathcal{F}_{n,g,\ell} &=& \frac{1}{24}\sum_{\nu = 1}^{n-1}\sum_{\gamma=0}^g \sum_{\lambda=0}^{\ell} \nu(n-\nu)\mathcal{F}_{\nu,\gamma,\lambda}\mathcal{F}_{n-\nu,g-\gamma,\ell-\lambda}\nonumber\\
&+& \frac{n(n+1)}{24}\mathcal{F}_{n,g-1,\ell}-\frac{n}{24}\mathcal{F}_{n,g,\ell-1}.\label{eq:modanomalyF}
\end{eqnarray}
These equations enable us to fix the $E_2$-dependent part of refined free energy as long as the datum of the lower order refined free energy are known. While the $E_2$-independent part can be determined by the fact that the space holomorphic modular form are finitely generated by $E_4$ and $E_6$. This method is effective to compute the refined free energy of low genus. As genus increases, the number of conditions grows faster than the number of coefficients that need to be fixed. For example, we can use this method to calculate refined free energy of three strings up to genus around 20.\\
\indent By the same spirit, we can compute the refined free energy for massive case, in which $\vec{m}_{E_8}$ are nonzero. The free energy here can be written as polynomials in nine Weyl invariant $E_8$ Jacobi forms $ A_1, A_2, A_3,A_4, A_5, B_2,B_3, B_4, B_6$, see the appendix of \cite{Sakai:2011xg}\cite{Huang:2013yta} for the explicit expressions. The subscripts of $A_i,B_i$ indicates the level of the characters of affine $E_8$ Lie algebra, and level $n$ characters contribute to $n$-string free energy. For example, in the case of two strings, the free energy can be written as the polynomials of $A_1^2,A_2,B_2$. In our case of three strings, the free energy can be written as a linear combinations of $A_1^3,A_1A_2,A_1B_2,A_3,B_3$. Besides, $A_i$ and $B_i$ are weight-four and weight-six Jacobi forms respectively. In the limit $\vec{m}\rightarrow0$, $A_i$ reduce to the Eisenstein series $E_4$, $B_i$ reduce to the Eisenstein series $E_6$, and the massive case reduces to massless case.\\
\indent We list some low genus refined free energy computed from topological string in the following:\footnote{These datum are computed in the notation of \cite{Huang:2013yta}, which coincide with the free energy computed from the domain method in Section three up to a factor $(-4)^{(g+l)}$. We keep this difference here because the datum here agree with the GV invariants computed in \cite{Huang:2013yta}, while the notation in the rest of the paper agree with \cite{Haghighat:2014pva} and the resulting coefficients allow the match $3E+3E\rightarrow 3H$ straightly.}
\begin{displaymath}
\begin{split}
F^{(0,0,3)}\frac{\eta^{36}}{q^{3/2}} =&\frac{1}{15552} \Big[54 E_2^2 A_1^3-54 E_4 A_1^3+135 E_4^2 A_1 A_2+28 E_4^3 A_3+27 E_2 A_1 (3 E_6 A_2+5 E_4 B_2)\\
&+E_6 (-28 E_6 A_3+225 A_1 B_2)\Big]
\end{split}
\end{displaymath}
\begin{displaymath}
\begin{split}
F^{(0,1,3)}\frac{\eta^{36}}{q^{3/2}}  =& \frac{1}{62208}\Big[78 E_2^3 A_1^3+45 E_2^2 A_1 (3 E_6 A_2+5 E_4 B_2)+E_2 (-54 E_4 A_1^3+297 E_4^2 A_1 A_2\\&+56 E_4^3 A_3
+E_6 (-56 E_6 A_3+495 A_1 B_2))
+8 (-3 E_6 (A_1^3-9 E_4 A_1 A_2)\\
&+10 E_6^2 B_3+5 E_4^2 (9 A_1 B_2-2 E_4 B_3))\Big]
\end{split}
\end{displaymath}
\begin{displaymath}
\begin{split}
F^{(1,0,3)}\frac{\eta^{36}}{q^{3/2}}  =& \frac{1}{124416}\Big[-54 E_2^3 A_1^3-27 E_2^2 A_1 (3 E_6 A_2+5 E_4 B_2)-E_2 (18 E_4 A_1^3+189 E_4^2 A_1 A_2\\
&+28 E_4^3 A_3+7 E_6 (-4 E_6 A_3+45 A_1 B_2))-2 (-36 E_6 A_1^3+189 E_4 E_6 A_1 A_2\\
&+315 E_4^2 A_1 B_2-80 E_4^3 B_3+80 E_6^2 B_3)\Big]
\end{split}
\end{displaymath}
\begin{displaymath}
\begin{split}
F^{(0,2,3)}\frac{\eta^{36}}{q^{3/2}} =& \frac{1}{2488320}\Big[575 E_2^4 A_1^3-141 E_4^2 A_1^3+1980 E_4^3 A_1 A_2+876 E_6^2 A_1 A_2+392 E_4^4 A_3\\
&+380 E_2^3 A_1 (3 E_6 A_2+5 E_4 B_2)-56 E_4 E_6 (7 E_6 A_3-85 A_1 B_2)+2 E_2^2 (-81 E_4 A_1^3\\
&+1575 E_4^2 A_1 A_2+280 E_4^3 A_3+35 E_6 (-8 E_6 A_3+75 A_1 B_2))
+2 E_2 (-136 E_6 A_1^3\\
&+2259 E_4 E_6 A_1 A_2+3765 E_4^2 A_1 B_2-800 E_4^3 B_3+800 E_6^2 B_3)\Big]
\end{split}
\end{displaymath}
\begin{displaymath}
\begin{split}
F^{(1,1,3)}\frac{\eta^{36}}{q^{3/2}}  =& \frac{1}{2488320}\Big[-390 E_2^4 A_1^3-225 E_2^3 A_1 (3 E_6 A_2+5 E_4 B_2)-E_2^2 (366 E_4 A_1^3\\
&+1935 E_4^2 A_1 A_2+280 E_4^3 A_3+5 E_6 (-56 E_6 A_3+645 A_1 B_2))-2 (-186 E_4^2 A_1^3\\
&+1935 E_4^3 A_1 A_2+936 E_6^2 A_1 A_2+392 E_4^4 A_3+E_4 E_6 (-392 E_6 A_3+4785 A_1 B_2))\\
&-8 E_2 (E_6 (-48 A_1^3+657 E_4 A_1 A_2)+250 E_6^2 B_3+5 E_4^2 (219 A_1 B_2-50 E_4 B_3))\Big]
\end{split}
\end{displaymath}
\begin{displaymath}
\begin{split}
F^{(2,0,3)}\frac{\eta^{36}}{q^{3/2}} =& \frac{1}{9953280}\Big[270 E_2^4 A_1^3+135 E_2^3 A_1 (3 E_6 A_2+5 E_4 B_2)+E_2^2 (486 E_4 A_1^3\\
&+1215 E_4^2 A_1 A_2+140 E_4^3 A_3+5 E_6 (-28 E_6 A_3+405 A_1 B_2))+2 (-306 E_4^2 A_1^3\\
&+2565 E_4^3 A_1 A_2+1296 E_6^2 A_1 A_2+532 E_4^4 A_3+E_4 E_6 (-532 E_6 A_3+6435 A_1 B_2))\\
&+2 E_2 (-72 E_6 A_1^3+2133 E_4 E_6 A_1 A_2+3555 E_4^2 A_1 B_2-800 E_4^3 B_3+800 E_6^2 B_3)\Big]
\end{split}
\end{displaymath}
\subsection{Domain wall method for two E-strings}
Recently, a new approach based on the computation of M9 domain walls was proposed, which in principle may enable us to obtain the explicit expressions for the elliptic genus of $n$ E-strings \cite{Haghighat:2014pva}. This method is supposed to provide an all-genus amplitude, which is impossible to achieve with the technique of topological string theory so far. This method also provides a new perspective to look at $n$ heterotic strings. The M9 domain wall blocks can be used to construct a new formula for the elliptic genus of heterotic strings, which has completely different appearance from the orbifold formula.\\
\indent The basic idea of this approach is to view the theory of $n$ M2-branes on $\mathbb{R}\times T^2$ as a quantum mechanical system on $\mathbb{R}$. This is reasonable because the elliptic genus does not depend on the size of the torus but only its complex structure. Under this reduction the states in the Hilbert space of $n$ M2- branes are labelled by Young diagrams of level $n$ \cite{Haghighat:2013gba,Kim:2010mr}. On the other hand, the M5-branes and M9-branes can be regarded as operators or states in this quantum mechanical system. They are called M5 domain walls and M9 domain walls respectively. Based on earlier works \cite{Iqbal:2012, Lockhart:2012}, the explicit formula for M5 domain walls was obtained in \cite{Haghighat:2013gba}, by relating M-strings to certain toric Calabi-Yau manifolds and using the (refined) topological vertex \cite{AKMV, IKV} to compute the corresponding partition functions. This allows one to compute the elliptic genus of $n$ M-strings. To deal with E-strings and H-strings, one also need to know the expression of M9 domain walls.\\
\indent Based on the known results of M5 domain walls and modular anomaly equation for E-strings, the form of M9 domain walls can be very restrictive. In \cite{Haghighat:2014pva}, the M9 domain walls for level one and level two Young diagrams are determined. The domain wall for $\nu=\ydiagram{1}$ can be obtained easily by the known expression of elliptic genus of one E-string. It is also easy to show that the elliptic genus of one heterotic string computed by combining the left and right M9 domain walls coincides with the well-known result. If this does not look so nontrivial, the two-string case does. The M9 domain walls for  $\nu=\ydiagram{2}$ and $\nu=\ydiagram{1,1}$  cannot be determined just by the form of M5 domain walls and modular anomaly equation. However, thanks to the structure of the space of weak Jacobi forms, the finite number of coefficients in M9 domain walls can be eventually determined by matching the ansatz of elliptic genus with the known results of low order refined free energy. Once the coefficients are fixed, the domain wall blocks are supposed to give an exact formula for the elliptic genus of two E-strings and we can check it by matching with the free energy datum from topological string to high orders. In the following, we briefly summarize this approach and results for two E-strings, see \cite{Haghighat:2014pva} for more details.\\
\indent One of the main ingredients of this approach is M5 domain wall blocks. To state the expression of M5 domain walls, it is convenient to introduce the notation
\begin{equation}
	\xi_+(\tau;z) = \prod_{k\geq 1} (1-Q_{\tau}^k e^{2\pi i z}), \quad \xi_-(\tau;z) = \prod_{k \geq 1}(1-Q_{\tau}^{k-1} e^{-2\pi i z}).
\end{equation}
These two functions can combine nicely into a theta function
\begin{equation}
	-i e^{-i \pi z} e^{\frac{\pi i \tau}{6}}\eta(\tau) \xi_-(\tau;z) \xi_+(\tau;z) = \theta_1(\tau;z).
\end{equation}
Though the general expression of M5 domain wall blocks has been found, we only write here the special cases which is relevant to our present use:
\begin{eqnarray}
	D^{\textrm{M5}}_{\emptyset \nu} & = & \prod_{(i,j)\in\nu}\frac{\theta_1(\tau;-m+\epsilon_1(\nu_i - j +1/2) - \epsilon_2(-i+1/2))\eta(\tau)^{-1}}{\xi_-(\tau;\epsilon_1(\nu_i-j) - \epsilon_2(\nu^t_j - i +1))\xi_+(\tau;\epsilon_1(\nu_i-j+1) - \epsilon_2(\nu^t_j-i))},\hspace{.3in} \\
	D^{\textrm{M5}}_{\nu \emptyset} & = & \prod_{(i,j)\in\nu}\frac{\theta_1(\tau;-m-\epsilon_1(\nu_i-j+1/2)+\epsilon_2(-i+1/2))\eta(\tau)^{-1}}{\xi_-(\tau;\epsilon_1(\nu_i-j+1)-\epsilon_2(\nu^t_j-i))\xi_+(\tau;\epsilon_1(\nu_i-j)-\epsilon_2(\nu^t_j-i+1))}.\hspace{.3in}\label{eq:DM5}
\end{eqnarray}
Note that $D^{\textrm{M5}}_{\emptyset \nu}$ and $D^{\textrm{M5}}_{\nu \emptyset}$ get exchanged under the map
\begin{equation}
	m \mapsto -m, \quad \xi_{\pm} \mapsto \xi_{\mp}.\label{eq:leftright}
\end{equation}
Since E-strings are realized by by M2-branes stretched between M5-brane and M9-brane, we can expect the elliptic genus of $n$ E-strings has the following form:
\begin{equation}
Z_n^{\textrm{E-str}} = \sum_{\vert\nu\vert = n} D^{\textrm{M9, L}}_{\nu} D^{\textrm{M5}}_{\nu \emptyset}.\label{eq:Estringdomain}
\end{equation}
For example,
\begin{equation}
	Z_1^{\textrm{E-str}} = D^{\textrm{M9, L}}_{\ydiagram{1}} D^{\textrm{M5}}_{\ydiagram{1}~ \emptyset},
\end{equation}
\begin{equation}
	Z_2^{\textrm{E-str}} = D^{\textrm{M9, L}}_{\ydiagram{2}} D^{\textrm{M5}}_{\ydiagram{2}~ \emptyset} + D^{\textrm{M9, L}}_{\ydiagram{1,1}} D^{\textrm{M5}}_{\ydiagram{1,1}~ \emptyset}.
\end{equation}
To determine the form of $D^{\textrm{M9}}_{\nu}$, one should obey several requests. First, the elliptic genus should transform with modular weight zero. Second, the elliptic genus of E-strings does not depend on the mass $ m = (\epsilon_4-\epsilon_3)/2 $. Third, The free energy of $n$ E-strings should be written as a level $n$ polynomial of Weyl invariants $A_i,B_i$. Forth, the elliptic genus should be symmetric under exchange of $ \epsilon_1 $ and $ \epsilon_2 $.\\
\indent The above constrains lead to the following ansatz for left M9 domain wall:
\begin{equation}
D^{M_9, L}_{\nu} = \frac{N^L_{\nu}(\tau; \vec{m}_{E_8,L},\epsilon_1,\epsilon_2)}{\eta(\tau)^{8|\nu|} B^L_\nu(\tau;\epsilon_1,\epsilon_2)F^R_\nu(\tau;\epsilon_1,\epsilon_2,m)},\label{eq:M9ansatz}
\end{equation}
where
\begin{equation}
B^L_\nu(\tau;\epsilon_1,\epsilon_2) = \prod_{(i,j)\in\nu}\xi_+(\epsilon_1(\nu_i-j+1)-\epsilon_2(\nu^t_j-i))\xi_-(\epsilon_1(\nu_i-j)-\epsilon_2(\nu^t_j-i+1))
\end{equation}
and
\begin{equation}
F^R_\nu(\tau;\epsilon_1,\epsilon_2) = \prod_{(i,j)\in\nu}\theta_1(-m-\epsilon_1(\nu_i-j+1/2)+\epsilon_2(-i+1/2))/\eta.
\end{equation}
These arrangements make the factors combine correctly with the M5 domain wall $ D^{M5}_{\nu\emptyset} $ (Equation \eqref{eq:DM5}).\\
\indent Likewise, the right M9 domain wall is given by
\begin{equation}
 D^{M_9, R}_{\nu} = \frac{N^R_{\nu}(\tau;\vec{m}_{E_8,R},\epsilon_1,\epsilon_2)}{\eta(\tau)^{8|\nu|} B^R_\nu(\tau;\epsilon_1,\epsilon_2)F^L_\nu(\tau;\epsilon_1,\epsilon_2,m)},\label{eq:M9Ransatz}\end{equation}
where
\begin{equation}
B^R_\nu(\tau;\epsilon_1,\epsilon_2) = \prod_{(i,j)\in\nu}\xi_-(\tau;\epsilon_1(\nu_i-j+1)-\epsilon_2(\nu^t_j-i))\xi_+(\tau;\epsilon_1(\nu_i-j)-\epsilon_2(\nu^t_j-i+1))
\end{equation}
and
\begin{equation}
F^L_\nu(\tau;\epsilon_1,\epsilon_2) = \prod_{(i,j)\in\nu}\theta_1(\tau;-m+\epsilon_1(\nu_i-j+1/2)-\epsilon_2(-i+1/2))/\eta(\tau).
\end{equation}
The transformation that exchanges left and right M5 domain walls leaves $ \epsilon_1 $ and $ \epsilon_2 $ fixed. Therefore,
\begin{equation}
N^R_{\nu}(\vec{m}_{E_8,R},\epsilon_1,\epsilon_2) = N^L_{\nu}(\vec{m}_{E_8,R},\epsilon_1,\epsilon_2) = N(\vec{m}_{E_8,R},\epsilon_1,\epsilon_2),
\end{equation}
The only difference between the numerators of the left and right domain walls is that they depend on the $m_i$ corresponding to different $ E_8 $ groups.\\
\indent To determine the explicit expression of the numerators, we need to use the modular anomaly equation. In \cite{Haghighat:2014pva}, an e-string holomorphic anomaly equation for each summand $ Z_\nu =D_\nu^{M9}D_{\nu\emptyset}^{M5} $ in \eqref{eq:Estringdomain} is proposed:
\begin{equation}
\partial_{E_2} Z_\nu^{E-str} = -\frac{1}{24}\Bigg[\epsilon_1\epsilon_2(|\nu|^2+|\nu|)-\epsilon_+^2|\nu|+|\nu|\bigg(\sum_{i=1}^8 m^2_{E_8,i}\bigg)\Bigg]\cdot Z_\nu^{E-str},\label{eq:EstrAnomaly}
\end{equation}
where $\epsilon_+=\epsilon_1+\epsilon_2$. Since this equation give the same modular anomaly for all Young diagram $v$ with $|\nu|=n$, it is easy to see
\begin{equation}
\partial_{E_2} Z_n^{E-str} = -\frac{1}{24}\Bigg[(n^2+n)\epsilon_1\epsilon_2-n(\epsilon_1+\epsilon_2)^2+n\bigg(\sum_{i=1}^8 m^2_{E_8,i}\bigg)\Bigg]\cdot Z_\nu^{E-str}.\label{eq:nEstrAnomaly}
\end{equation}
This anomaly equation can be derived from the modular form equation of E-string refined free energy \eqref{eq:modanomalyF}. It can also be derived from the modular equations for heterotic and M-strings by using the above ansatz of M9 domain walls.\\
\indent After the general consideration for $n$ E-strings, we now focus on the case of two E-strings. The two E-string elliptic genus can be written as
\begin{equation}
Z_2^{\textrm{E-str}} =  D_{\substack{\ydiagram{2}}}^{M9,L} D^{M5}_{\ydiagram{2}~\emptyset} +D_{\ydiagram{1,1}}^{M9,L}   D^{M5}_{\ydiagram{1,1}~\emptyset},
\end{equation}
where
\begin{equation}
D^{M5}_{\ydiagram{2}~\emptyset}= \frac{\theta_1(m+\epsilon_+/2)\eta^{-1}}{\xi_-(\epsilon_1)\xi_+(-\epsilon_2)}\frac{\theta_1(m+\epsilon_+/2+\epsilon_1)\eta^{-1}}{\xi_-(2\epsilon_1)\xi_+(\epsilon_1-\epsilon_2)}
\end{equation}
and
\begin{equation}
D^{M5}_{\ydiagram{1,1}~\emptyset}=\frac{\theta_1(m-\epsilon_+/2)\eta^{-1}}{\xi_+(\epsilon_1)\xi_-(-\epsilon_2)}\frac{\theta_1(m-\epsilon_+/2-\epsilon_2)\eta^{-1}}{\xi_+(-2\epsilon_2)\xi_-(\epsilon_1-\epsilon_2)}.
\end{equation}
This leads to the following ansatz for the M9 domain walls:
\begin{align}
D_{\ydiagram{2}}^{M9,L} &= \frac{N_{\ydiagram{2}}(\vec{m}_{E_8,L},\epsilon_1,\epsilon_2)/\eta^{16}}{\xi_+(\epsilon_1)\xi_-(-\epsilon_2)\xi_-(\epsilon_1-\epsilon_2)\xi_+(2\epsilon_1)\theta_1(m+\epsilon_+/2)\theta_1(m+\epsilon_+/2+\epsilon_1)\eta^{-2}},\\
D_{\ydiagram{1,1}}^{M9,L} &=  \frac{N_{\ydiagram{1,1}}(\vec{m}_{E_8, L},\epsilon_1,\epsilon_2)/\eta^{16}}{\xi_-(\epsilon_1)\xi_+(-\epsilon_2)\xi_+(\epsilon_1-\epsilon_2)\xi_-(-2\epsilon_2)\theta_1(m-\epsilon_+/2)\theta_1(m-\epsilon_+/2-\epsilon_2)\eta^{-2}}.
\end{align}
Then the two E-string elliptic genus can be written as
\begin{equation}
Z_2^{\textrm{E-str}} = -\frac{N_{\ydiagram{2}}(\vec{m}_{E_8,L},\epsilon_1,\epsilon_2)/\eta^{16}}{\theta_1(\epsilon_1)\theta_1(\epsilon_2)
\theta_1(\epsilon_1-\epsilon_2)\theta_1(2\epsilon_1)\eta^{-4}}-\frac{N_{\ydiagram{1,1}}(\vec{m}_{E_8,L},
\epsilon_1,\epsilon_2)/\eta^{16}}{\theta_1(\epsilon_1)\theta_1(\epsilon_2)\theta_1(\epsilon_2-\epsilon_1)
\theta_1(2\epsilon_2)\eta^{-4}}.\label{eq:Z2Ansatz}
\end{equation}
The modular weight of $ N_{\ydiagram{2}} $ and $ N_{\ydiagram{1,1}}$ should be 8 to guarantee the modular invariance of $Z_2^{\textrm{E-str}}$. Besides, $ N_{\ydiagram{2}} $ and $ N_{\ydiagram{1,1}}$ should be written as linear combinations of the three level-two Weyl invariant $E_8$ Jacobi forms $ A_1^2, A_2,$ and $ B_2 $. To determine the explicit expression of $N$, we need to exploit the modular anomaly equation. From Eq \eqref{eq:EstrAnomaly} and \eqref{eq:Z2Ansatz}, one obtains
\begin{equation}
\partial_{E_2}\log N_{\ydiagram{2}} = -\frac{1}{24}\left[4\epsilon_1^2+\left(\sum_{i=1}^8(m_{E_8,i}^L)^2\right)\right].
\end{equation}
This means $ N_{\ydiagram{2}} $ is a function of $ \epsilon_1$ and not $ \epsilon_2 $, and it transforms with index 2 with respect to $ \epsilon_1 $ under modular transformation. Once $ N_{\ydiagram{2}} $ is determined, $ N_{\ydiagram{1,1}} $ is determined at the same time, since $N_{\ydiagram{2}}(\tau;\vec{m}_{E_8,L},\epsilon_2,\epsilon_1) =  N_{\ydiagram{1,1}}(\tau;\vec{m}_{E_8,L},\epsilon_1,\epsilon_2)$.\\
$N_{\ydiagram{2}}(\tau;\vec{m}_{E_8,L},\epsilon_1,\epsilon_2)$ does not contain $\epsilon_2$, this allows us to use the known mathematical results for Jacobi forms to uniquely fix the expression of domain walls. For weak Jacobi forms, there are the following structure theorem: \cite{Zagierbook}\cite{Dabholkar:2012nd}\\\newline
{ \noindent\emph{The weak Jacobi forms with modular parameter $ \tau $ and elliptic parameter $ \epsilon $ of index $ k $ and even weight $ w $ form a polynomial ring which is generated by the four modular forms $ E_4(\tau), E_6(\tau), \phi_{0,1}(\epsilon,\tau), $ and $\phi_{-2,1}(\epsilon,\tau) $, where}
\[ \phi_{-2,1}(\epsilon,\tau) = -\frac{\theta_1(\epsilon;\tau)^2}{\eta^6(\tau)}\qquad \text{\emph{and}}\qquad
\phi_{0,1}(\epsilon,\tau) = 4\left[\frac{\theta_2(\epsilon;\tau)^2}{\theta_2(0;\tau)^2}+\frac{\theta_3(\epsilon;\tau)^2}{\theta_3(0;\tau)^2}+\frac{\theta_4(\epsilon;\tau)^2}{\theta_4(0;\tau)^2}\right]\]
\emph{are Jacobi forms of index 1, respectively of weight $ -2 $ and 0.}}\newline\\
\indent By matching against the known free energy calculated from topological strings, the explicit expression of domain walls can be determined as
\begin{align}
&N_{\ydiagram{2}}(\vec{m}_{E_8,L},\epsilon_1)= \frac{1}{576}\bigg[4 A_1^2(\phi_{0,1}(\epsilon_1)^2- E_4\phi_{-2,1}(\epsilon_1)^2)\hspace{2in}\nonumber\\
&\hspace{.1in}+3A_2(E_4^2\phi_{-2,1}(\epsilon_1)^2-E_6\phi_{-2,1}(\epsilon_1)\phi_{0,1}(\epsilon_1))+5 B_2(E_6\phi_{-2,1}(\epsilon_1)^2-E_4 \phi_{-2,1}(\epsilon_1)\phi_{0,1}(\epsilon_1))\bigg],
\end{align}
The coefficients in the above formula are fixed by the datum of refined free energy up to genus two, which means $g+l=2$. Since our ansatz for the elliptic genus of two E-strings is supposed to be exact, we can further compute the refined free energy to all genus. We check this up to genus 10 and the match is perfect.\\
\indent The above M9 domain walls can also be used to compute the two heterotic string partition function. Since heterotic strings are realized by the M2-branes stretched between two M9-branes, it is natural to expect the H-strings amplitude can obtained by combining the left and right M9 domain wall blocks. Considering also the symmetry among $\epsilon_{1,2,3,4}$, a new formula for the elliptic genus of two heterotic strings was proposed in \cite{Haghighat:2014pva}:
\begin{equation} \label{eq:Hetstrres}
Z_2^{\textrm{het}} = D_{\ydiagram{2}}^{M9,L}(\vec{m}_{E_8,L})D_{\ydiagram{2}}^{M9,R}(\vec{m}_{E_8,R})+(\epsilon_1\leftrightarrow\epsilon_2)+(\epsilon_1\leftrightarrow\epsilon_3)+(\epsilon_1\leftrightarrow\epsilon_4).
\end{equation}
in which
\begin{eqnarray}
~ & ~ & D_{\ydiagram{2}}^{M9,L}(\vec{m}_{E_8,L})D_{\ydiagram{2}}^{M9,R}(\vec{m}_{E_8,R}) \nonumber \\
~ & = & -\frac{N_{\ydiagram{2}} (\vec{m}_{E_8,L},\epsilon_1)N_{\ydiagram{2}} (\vec{m}_{E_8,R},\epsilon_1)}{\eta(\tau)^{24}\theta_1(\epsilon_1)\theta_1(\epsilon_2)\theta_1(\epsilon_3)\theta_1(\epsilon_4)\theta_1(2\epsilon_1)\theta_1(\epsilon_1-\epsilon_2)\theta_1(\epsilon_1-\epsilon_3)\theta_1(\epsilon_1-\epsilon_4)}\nonumber\\
~ & = & -\frac{N_{\ydiagram{2}} (\vec{m}_{E_8,L},\epsilon_1)N_{\ydiagram{2}} (\vec{m}_{E_8,R},\epsilon_1)}{\eta(\tau)^{24}\theta_1(\epsilon_1)\theta_1(\epsilon_2)\theta_1(\epsilon_3)\theta_1(\epsilon_4)\theta_1(2\epsilon_1)\theta_1(\epsilon_1-\epsilon_2)\theta_1(\epsilon_1-\epsilon_3)\theta_1(\epsilon_1-\epsilon_4)}
\end{eqnarray}
On the other hand, the orbit formula shows the elliptic genus of two heterotic strings is
\begin{eqnarray}
	Z_2^{\textrm{het}}(\tau,\vec{\epsilon}, \vec{m}_{E_8\times E_8})& = & \frac{1}{2}\left[\left(Z_1^{\textrm{het}}(\tau,\vec{\epsilon},\vec{m}_{E_8\times E_8})\right)^2 + Z_1^{\textrm{het}}(2\tau,2\vec{\epsilon},2\vec{m}_{E_8\times E_8})\right. \nonumber \\
	~ & ~ & + \left. Z_1^{\textrm{het}}(\frac{\tau}{2},\vec{\epsilon},\vec{m}_{E_8\times E_8})+Z_1^{\textrm{het}}(\frac{\tau+1}{2},\vec{\epsilon},\vec{m}_{E_8\times E_8})\right]. \nonumber \\ \label{eq:symhet2}
\end{eqnarray}
In \cite{Haghighat:2014pva}, these two formulas are checked and match up to powers of $Q_{\tau}^8$ with a generic choice of $E_8\times E_8$ Wilson lines.
\section{Elliptic genus of three E-strings and three H-strings}
In this section, we will calculate the M9 domain wall blocks of level three partition and use these blocks to construct the formulas for the elliptic genus of three E-strings and H-strings. Specializing the general ansatz for $n$ E-strings (Eq \eqref{eq:Estringdomain}) to the case of three, we obtain the following formula for the elliptic genus of three E-strings:
\begin{equation}
Z_3^{\textrm{E-str}} =  D_{\substack{\ydiagram{3}}}^{M9,L} D^{M5}_{\ydiagram{3}~\emptyset} +D_{\ydiagram{2,1}}^{M9,L}   D^{M5}_{\ydiagram{2,1}~\emptyset}
+D_{\ydiagram{1,1,1}}^{M9,L}   D^{M5}_{\ydiagram{1,1,1}~\emptyset},
\end{equation}
Based on the known results of M5 domain walls and the ansatz of M9 domain walls in Section 2.3, this formula can be explicitly written as
\begin{equation}
\begin{split}
Z_3^{\mathrm{E-str}}=&\frac{N_{\ydiagram{3}}(\epsilon_1)}{\eta^{18}\theta_1(3\epsilon_1)\theta_1(2\epsilon_1-\epsilon_2)
\theta_1(2\epsilon_1)\theta_1(\epsilon_1-\epsilon_2)\theta_1(\epsilon_1)\theta_1(-\epsilon_2)}\\
+&\frac{N_{\ydiagram{2,1}}(\epsilon_1,\epsilon_2)}{\eta^{18}\theta_1(2\epsilon_1-\epsilon_2)
\theta_1(\epsilon_1-2\epsilon_2)\theta_1(\epsilon_1)^2\theta_1(-\epsilon_2)^2}\\
+&\frac{N_{\ydiagram{1,1,1}}(\epsilon_2)}{\eta^{18}\theta_1(\epsilon_1-2\epsilon_2)\theta_1(-3\epsilon_2)
\theta_1(\epsilon_1-\epsilon_2)\theta_1(-2\epsilon_2)\theta_1(\epsilon_1)\theta_1(-\epsilon_2)}.
\end{split}
\end{equation}
The main task of our paper is to determine the numerators $N_{\ydiagram{3}}$, $N_{\ydiagram{2,1}}$ and $N_{\ydiagram{1,1,1}}$.\\
\indent In the case of three strings, there are some new phenomena arising, which make it more difficult to handle than the case of two strings. From the modular anomaly equation, we find the numerator of the domain wall for Young diagram $\nu=\ydiagram{2,1}$ has non-diagonal index matrix, which means it must be a function of $\epsilon_1$ and $\epsilon_2$ simultaneously. This is quite different from the situation of two strings where the numerator $N_{\ydiagram{2}}$ just contains $\epsilon_1$ but no $\epsilon_2$ and $N_{\ydiagram{1,1}}$ just contains $\epsilon_2$ but no $\epsilon_1$. The structure of the space of weak Jacobi forms with one elliptic parameter enables us to expand the numerators as the polynomials of $A_i,B_i$ and the known bases of weak Jacobi forms. However, since $N_{\ydiagram{2,1}}$ contain both $\epsilon_1$ and $\epsilon_2$, we need to know the structure of the space of Jacobi forms of higher degrees. So far little is known for these objects, see \cite{Ziegler}\cite{Yang} for some basic knowledge. Lack of bases for the space of weak Jacobi form with general index quadratic form makes it impossible to directly follow the approach in the two-string case. In this section, we propose an ansatz by transforming the index matrix to a diagonal one and make the problem accessible by the known knowledge of the bases of weak Jacobi form with one elliptic parameter. By matching our formula with the free energy computed from the topological string, we find the infinite free energy datum are not sufficient to fix all coefficients in our domain wall ansatz, though the resulting elliptic genera of three E-strings and three H-strings are already fixed and exact to arbitrarily high orders. This completes the endeavor to search for an explicit all-genus expression for the elliptic genus of three E-strings. Besides, the requirement for the terms proportional to the free parameters to vanish results in an identity among Jacobi forms. Based on some consideration on simplification, we can also fix the domain wall blocks in massless case by properly choosing the unfixed parameter, since the expansion coefficients of domain walls are supposed to have physical interpretation themselves.\\
\indent This section is divided as follows: in Section 3.1 we calculate the modular anomaly of the elliptic genus of three E-strings and the M9 domain walls for the level three partition; in Section 3.2 we propose our explicit ansatz for the M9 domain walls; in Section 3.3 we use our domain wall ansatz to express the elliptic genus of three E-strings and match it against the refined free energy computed from topological string, we show that the relations obtained from comparison are not sufficient to fix all coefficients, but one identity among Jacobi forms guarantee our formula for elliptic genus indeed contain no free parameters; in Section 3.4, based on our M9 domain wall blocks, we propose a novel expression for the elliptic genus of three heterotic strings; in Section 3.5, we make some additional comments on the domain wall expressions we obtained.
\subsection{E-string holomorphic anomaly}
From the modular anomaly equation for $Z_{\nu}^{E-str}$ one obtains for $\nu=\ydiagram{3},\ \ydiagram{2,1}\ \ \mathrm{or}\ \ \ydiagram{1,1,1}\ $,
\begin{equation}
\partial_{E_2}\log Z_\nu^{E-str} = -\frac{1}{24}\Bigg[-3(\epsilon_1-\epsilon_2)^2+3\bigg(\sum_{i=1}^8 m^2_{E_8,i}\bigg)\Bigg],\label{eq:3EstrAnomaly}
\end{equation}
Note Jacobi theta function $\theta_1$ has index 1/2, which means it has the following modular anomaly
\begin{equation}
\partial_{E_2}\log \theta_1(\epsilon_1)=-\frac{1}{24}.
\end{equation}
Together with Eq \eqref{eq:Estringdomain}, it is easy to obtain the modular anomaly for the numerators of M9 domain walls:
\begin{equation}
\partial_{E_2}\log N_{\ydiagram{3}}(\epsilon_1,\epsilon_2)=-\frac{1}{24}\Big(16\epsilon_1^2+3\sum m_i^2\Big)
\end{equation}
\begin{equation}
\partial_{E_2}\log N_{\ydiagram{2,1}}(\epsilon_1,\epsilon_2)=
-\frac{1}{24}\Big(4\epsilon_1^2-2\epsilon_1\epsilon_2+4\epsilon_2^2+3\sum m_i^2\Big)
\end{equation}
\begin{equation}
\partial_{E_2}\log N_{\ydiagram{1,1,1}}(\epsilon_1,\epsilon_2)=-\frac{1}{24}\Big(16\epsilon_2^2+3\sum m_i^2\Big)
\end{equation}
These show that $N_{\ydiagram{3}}$ transforms with index 8 with respect to $\epsilon_1$ and $N_{\ydiagram{1,1,1}}$ transforms with index 8 with respect to $\epsilon_2$. However, $N_{\ydiagram{2,1}}$ transforms with a non-diagonal index matrix. This is the main obstruction since we do not know the precise bases of weak Jacobi forms with generic index matrix. In the following subsection we will show that the key point for $ N_{\ydiagram{2,1}}$ is to transform its index matrix to a diagonal one in the following way:
\begin{equation}
2\epsilon_1^2-\epsilon_1\epsilon_2+2\epsilon_2^2=
5\Big(\frac{\epsilon_1-\epsilon_2}{2}\Big)^2+3\Big(\frac{\epsilon_1+\epsilon_2}{2}\Big)^2
\end{equation}
This transformation reduces the problem to a simpler one which we can handle with the known results about Jacobi forms.
\subsection{M9 domain walls}
To determine the explicit expression of the numerators, we need to recall the general requirements in section 2.3 and specialize to the case of three strings. Modular invariance of $Z_3^{E-str}$ requires the modular weight of $N_{\ydiagram{3}}$, $N_{\ydiagram{2,1}}$ and $N_{\ydiagram{1,1,1}}$ to be 12. The modular anomaly equations show that they can be written as linear combinations of five level-three Weyl[$E_8$]-invariant modular forms $A_1^3$, $A_1A_2$, $A_1B_2$, $A_3$ and $B_3$.\\
\indent These requirements force $N_{\ydiagram{3}}$ to have the following form:
\begin{equation}
N_{\ydiagram{3}}(\epsilon_1,\epsilon_2)=A_1^3f_1(\epsilon_1)
+A_1A_2f_2(\epsilon_1)+A_1B_2f_3(\epsilon_1)+A_3f_4(\epsilon_1)+B_3f_5(\epsilon_1),
\end{equation}
where $f_i(\epsilon_1)$ are Jacobi forms of index 8 with elliptic parameter $\epsilon_1$, respectively of modular weight $0,4,2,8,6$. Since $N_{\ydiagram{3}}(\epsilon_1,\epsilon_2)=N_{\ydiagram{1,1,1}}(\epsilon_2,\epsilon_1)$, we do not repeat the formulas for $N_{\ydiagram{1,1,1}}$ in the following.\\
\indent The formulas in the case of three strings are quite long due to the large indices. To save the space and make the formulas more clear, we introduce the following notations. Denote
\begin{equation}
\Phi(\epsilon_1)=\begin{pmatrix}
\phi_{0,1}(\epsilon_1)^8 & \phi_{0,1}(\epsilon_1)^7\phi_{-2,1}(\epsilon_1) & \cdots & \phi_{0,1}(\epsilon_1)\phi_{-2,1}(\epsilon_1)^7 & \phi_{-2,1}(\epsilon_1)^8
\end{pmatrix}
\end{equation}
as a row vector. Then $f_1(\epsilon_1)=\Phi(\epsilon_1)F_1$, where
\begin{equation}
F_1=\begin{pmatrix}
c_{11} \\
0 \\
c_{12} E_4 \\
c_{13} E_6 \\
c_{14} E_4^2 \\
c_{15} E_4E_6 \\
c_{16} E_4^3+c_{17} E_6^2 \\
c_{18} E_4^2E_6 \\
c_{19} E_4^4+c_{110}E_6^2E_4
\end{pmatrix}.
\end{equation}
This is because the polynomial ring of weak Jacobi forms is generated by $E_4$, $E_6$, $\phi_{0,1}$ and $\phi_{-2,1}$, as we have stated in Section 2.3. For the rest of $f_i(\epsilon_1)$, we all have $f_i(\epsilon_1)=\Phi(\epsilon_1)F_i$, where $F_i$ are some column vectors.
\begin{displaymath}
\begin{split}
&F_2=\begin{pmatrix}
c_{21}E_4\\
c_{22}E_6\\
c_{23}E_4^2\\
c_{24}E_4E_6\\
c_{25}E_4^3+c_{26}E_6^2\\
c_{27}E_4^2E_6\\
c_{28}E_4^4+c_{29}E_6^2E_4\\
c_{210}E_6^3+c_{211}E_4^3E_6\\
c_{212}E_4^2E_6^2+c_{213}E_4^5
\end{pmatrix};\quad F_3=\begin{pmatrix}
c_{31}E_4\\
 c_{32}E_6\\
 c_{33}E_4^2\\
 c_{34}E_4E_6\\
c_{35}E_4^3 + c_{36}E_6^2\\
c_{37}E_4^2E_6\\
c_{38}E_4^4 + c_{39}E_6^2E_4\\
c_{310}E_6^3 + c_{311}E_4^3E_6
\end{pmatrix};\\
\end{split}
\end{displaymath}
\begin{displaymath}
\begin{split}
&F_4=\begin{pmatrix}
c_{41}E_4^2\\
c_{42}E_4E_6\\
c_{43}E_4^3 + c_{44}E_6^2\\
c_{45}E_4^2E_6\\
c_{46}E_4^4 + c_{47}E_6^2E_4\\
c_{48}E_6^3 + c_{49}E_4^3E_6\\
c_{410}E_4^2E_6^2 + c_{411}E_4^5\\
c_{412}E_4E_6^3 + c_{413}E_4^4E_6\\
c_{414}E_4^6 + c_{415}E_4^3E_6^2 + c_{416}E_6^4\\
\end{pmatrix};\quad F_5=\begin{pmatrix}
c_{51}E_6\\
c_{52}E_4^2\\
 c_{53}E_4E_6\\
c_{54}E_4^3 + c_{55}E_6^2\\
c_{56}E_4^2E_6\\
c_{57}E_4^4 + c_{58}E_6^2E_4\\
c_{59}E_6^3 + c_{510}E_4^3E_6\\
c_{511}E_4^2E_6^2 + c_{512}E_4^5\\
c_{513}E_4E_6^3 + c_{514}E_4^4E_6
\end{pmatrix};
\end{split}
\end{displaymath}
In total, there are 64 parameters in $N_{\ydiagram{3}}$.\\
\indent Now we turn to $N_{\ydiagram{2,1}}$. As we already stated in the last subsection, to solve the problem with the non-diagonal index matrix, we must transform it to a diagonal one:
\begin{equation}
2\epsilon_1^2-\epsilon_1\epsilon_2+2\epsilon_2^2=
5\Big(\frac{\epsilon_1-\epsilon_2}{2}\Big)^2+3\Big(\frac{\epsilon_1+\epsilon_2}{2}\Big)^2
=5\omega_1^2+3\omega_2^2
\end{equation}
Of course the transformation is not unique. There are some good reasons why this ansatz could be right. For example, if the sum of the indices of two new variables is 8, then almost every term in $N_{\ydiagram{3}}$ and $N_{\ydiagram{1,1,1}}$ has its counterpart in $N_{\ydiagram{2,1}}$. As long as we have diagonal index matrix, we can deal with $\omega_1$ and $\omega_2$ separately. This bypasses the obstruction that the basis for Jacobi form of higher degree are still unknown. However, we still do not know this ansatz should be right a priori. Fortunately, the coincidence with the results from topological string shows this is indeed the right way.\\
\indent After the transformation, $N_{\ydiagram{2,1}}$ becomes a Jacobi form of index 5 with respect to $\omega_1$ and index 3 with respect to $\omega_2$. Together with the requirement on modular weight, it must have the following form:

\begin{equation}
N_{\ydiagram{2,1}}(\epsilon_1,\epsilon_2)=A_1^3h_1(\omega_1,\omega_2)
+A_1A_2h_2(\omega_1,\omega_2)+A_1B_2h_3(\omega_1,\omega_2)
+A_3h_4(\omega_1,\omega_2)+B_3h_5(\omega_1,\omega_2),
\end{equation}
where $h_i(\omega_1,\omega_2)$ are Jacobi forms of index 5 with elliptic parameter $\omega_1$ and index 3  with elliptic parameter $\omega_2$, respectively of modular weight $0,4,2,8,6$.\\
\indent According to the structure theorem of weak Jacobi form, we can expand $h_i(\omega_1,\omega_2)$ in the following way. Denote
\begin{equation}
\Psi_1(\omega_1)=\begin{pmatrix}
\phi_{0,1}(\omega_1)^5 & \phi_{0,1}(\omega_1)^4\phi_{-2,1}(\omega_1) & \cdots & \phi_{0,1}(\omega_1)\phi_{-2,1}(\omega_1)^4 & \phi_{-2,1}(\omega_1)^5
\end{pmatrix}
\end{equation}
as a row vector and
\begin{equation}
\Psi_2(\omega_2)=\begin{pmatrix}
\phi_{0,1}(\omega_2)^3 & \phi_{0,1}(\omega_2)^2\phi_{-2,1}(\omega_2) & \phi_{0,1}(\omega_2)\phi_{-2,1}(\omega_2)^2 & \phi_{-2,1}(\omega_2)^3
\end{pmatrix}
\end{equation}
as a column vector. Then $h_1(\omega_1,\omega_2)=\Psi_1(\omega_1)H_1\Psi_2(\omega_2)$, where
\begin{equation}
H_1=\begin{pmatrix}
a_{111} & 0 & a_{113}E_4 & a_{114}E_6 \\
0 & a_{122}E_4 & a_{123}E_6 & a_{124} E_4^2\\
a_{131}E_4 & a_{132}E_6 & a_{133}E_4^2 & a_{134}E_4E_6 \\
a_{141}E_6 & a_{142} E_4^2 & a_{143} E_4E_6 & a_{144} E_4^3+b_{144}E_6^2\\
a_{151}E_4^2 & a_{152}E_4E_6 & a_{153}E_4^3+b_{153}E_6^2 & a_{154}E_4^2E_6\\
a_{161}E_4E_6 & a_{162}E_4^3+b_{162}E_6^2 & a_{163} E_4^2E_6 & a_{164}E_4^4+b_{164}E_4E_6^2
\end{pmatrix}.
\end{equation}
For the rest of $h_i(\omega_1,\omega_2)$, we all have $h_i(\omega_1,\omega_2)=\Psi_1(\omega_1)H_i\Psi_2(\omega_2)$, where $H_i$ are some $4\times6$ matrices.
\begin{displaymath}
\begin{split}
H_2=&\\
&\begin{pmatrix}
a_{211}E_4 & a_{212}E_6 & a_{213}E_4^2 & a_{214}E_4E_6\\
 a_{221}E_6 & a_{222}E_4^2 & a_{223}E_4E_6 & a_{224}E_4^3 + b{224}E_6^2\\
 a_{231}E_4^2& a_{232}E_4E_6& a_{233}E_4^3 + b{233}E_6^2& a_{234}E_4^2E_6\\
 a_{241}E_4E_6& a_{242}E_4^3 + b_{242}E_6^2& a_{243}E_4^2E_6&
  a_{244}E_4^4 + b_{244}E_4E_6^2\\
 a_{251}E_4^3 + b_{251}E_6^2& a_{252}E_4^2E_6& a_{253}E_4^4 + b_{253}E_4E_6^2&
  a_{254}E_4^3E_6 + b_{254}E_6^3\\
 a_{261}E_4^2E_6& a_{262}E_4^4 + b_{262}E_4E_6^2& a_{263}E_4^3E_6 + b_{263}E_6^3&
  a_{264}E_4^5 + b_{264}E_4^2E_6^2
\end{pmatrix};\\
\end{split}
\end{displaymath}
\begin{displaymath}
\begin{split}
H_3=&\\
&\begin{pmatrix}
0& a_{312}E_4& a_{313}E_6& a_{314}E_4^2\\
 a_{321}E_4& a_{322}E_6& a_{323}E_4^2& a_{324}E_4E_6\\
 a_{331}E_6& a_{332}E_4^2& a_{333}E_4E_6& a_{334}E_4^3 + b_{334}E_6^2\\
 a_{341}E_4^2& a_{342}E_4E_6& a_{343}E_4^3 + b_{343}E_6^2& a_{344}E_4^2E_6\\
 a_{351}E_4E_6& a_{352}E_4^3 + b_{352}E_6^2& a_{353}E_4^2E_6&
  a_{354}E_4^4 + b_{354}E_4E_6^2\\
 a_{361}E_4^3 + b_{361}E_6^2& a_{362}E_4^2E_6& a_{363}E_4^4 + b_{363}E_4E_6^2&
  a_{364}E_4^3E_6 + b_{364}E_6^3
\end{pmatrix};\\
\end{split}
\end{displaymath}
\begin{displaymath}
\begin{split}
&H_4=\\
&\begin{pmatrix}
a_{411}E_4^2& a_{412}E_4E_6& a_{413}E_4^3 + b_{413}E_6^2& a_{414}E_4^2E_6\\
a_{421}E_4E_6& a_{422}E_4^3 + b_{422}E_6^2& a_{423}E_4^2E_6&  a_{424}E_4^4 + b_{424}E_4E_6^2\\
a_{431}E_4^3 + b_{431}E_6^2& a_{432}E_4^2E_6& a_{433}E_4^4 + b_{433}E_4E_6^2&  a_{434}E_6^3 + b_{434}E_4^3E_6\\
a_{441}E_4^2E_6& a_{442}E_4^4 + b{442}E_4E_6^2& a_{443}E_6^3 + b_{443}E_4^3E_6&  a_{444}E_4^2E_6^2 + b_{444}E_4^5\\
a_{451}E_4^4 + b_{451}E_4E_6^2& a_{452}E_6^3 + b_{452}E_4^3E_6&  a_{453}E_4^2E_6^2 + b_{453}E_4^5& a_{454}E_4E_6^3 + b_{454}E_4^4E_6\\
a_{461}E_6^3 + b_{461}E_4^3E_6& a_{462}E_4^2E_6^2 + b_{462}E_4^5&  a_{463}E_4E_6^3 + b_{463}E_4^4E_6& a_{464}E_4^6 + b_{464}E_4^3E_6^2 + d_{464}E_6^4
\end{pmatrix};\\
\end{split}
\end{displaymath}
\begin{displaymath}
\begin{split}
&H_5=\\
&\begin{pmatrix}
a_{511}E_6& a_{512}E_4^2& a_{513}E_4E_6& a_{514}E_4^3 + b_{514}E_6^2\\
a_{521}E_4^2& a_{522}E_4E_6& a_{523}E_4^3 + b_{523}E_6^2& a_{524}E_4^2E_6\\
a_{531}E_4E_6& a_{532}E_4^3 + b_{532}E_6^2& a_{533}E_4^2E_6&  a_{534}E_4^4 + b_{534}E_4E_6^2\\
a_{541}E_4^3 + b_{541}E_6^2& a_{542}E_4^2E_6& a_{543}E_4^4 + b_{543}E_4E_6^2&
a_{544}E_6^3 + b_{544}E_4^3E_6\\
a_{551}E_4^2E_6& a_{552}E_4^4 + b_{552}E_4E_6^2& a_{553}E_6^3 + b_{553}E_4^3E_6&
a_{554}E_4^2E_6^2 + b_{554}E_4^5\\
a_{561}E_4^4 + b_{561}E_4E_6^2& a_{562}E_6^3 + b_{562}E_4^3E_6&
a_{563}E_4^2E_6^2 + b_{563}E_4^5& a_{564}E_4E_6^3 + b_{564}E_4^4E_6
\end{pmatrix};\\
\end{split}
\end{displaymath}
In total, there are 170 parameters in $N_{\ydiagram{2,1}}$.\\
\indent We will fix these coefficients in the following subsection.
\subsection{Elliptic genus of three E-strings}
We already know the elliptic genus of three E-strings can be written as
\begin{equation}\label{eq:3Eellipticgenus}
\begin{split}
Z_3^{\mathrm{E-str}}=&\frac{N_{\ydiagram{3}}(\epsilon_1)}{\eta^{18}\theta_1(3\epsilon_1)\theta_1(2\epsilon_1-\epsilon_2)
\theta_1(2\epsilon_1)\theta_1(\epsilon_1-\epsilon_2)\theta_1(\epsilon_1)\theta_1(-\epsilon_2)}\\
+&\frac{N_{\ydiagram{2,1}}(\epsilon_1,\epsilon_2)}{\eta^{18}\theta_1(2\epsilon_1-\epsilon_2)
\theta_1(\epsilon_1-2\epsilon_1)\theta_1(\epsilon_1)^2\theta_1(-\epsilon_2)^2}\\
+&\frac{N_{\ydiagram{1,1,1}}(\epsilon_2)}{\eta^{18}\theta_1(\epsilon_1-2\epsilon_2)\theta_1(-3\epsilon_2)
\theta_1(\epsilon_1-\epsilon_2)\theta_1(-2\epsilon_2)\theta_1(\epsilon_1)\theta_1(-\epsilon_2)}.
\end{split}
\end{equation}
Since we have obtained the expression of the numerators of M9 domain wall blocks, the form of above formula is determined up to 234 parameters, in which 64 parameters come from $N_{\ydiagram{3}}$ and $N_{\ydiagram{1,1,1}}$ and 170 parameters come from $N_{\ydiagram{2,1}}$. Using $F_3=Z_3-Z_1Z_2+Z_1^3/3$, we can also obtain the total free energy $F_3$ with 234 parameters.\\
\indent To determine the  coefficients, we exploit the results of refined free energy computed from topological string on half K3 surface and match them against our ansatz. It is rather surprising that the infinite datum from topological string are not sufficient to uniquely fix all the coefficients in our expression. We obtain 230 relations from the match between refined free energy up to $g+l=6$. After that, even though there are 4 unfixed parameters in our expression, the free energy matches automatically from $g+l=7$ to $g+l=12$. On one hand, this indicates that our formula should be right since otherwise there must be many inconsistent equations other than redundant equations. On the other hand, it rules out the possibility to fix all coefficients by just comparing the refined free energy. This is quite different from the case of two E-strings. One possibility is that though the coefficients in $N_{\ydiagram{2,1}}$ are independent for $\omega_1$ and $\omega_2$, they are not completely independent for $\epsilon_1$ and $\epsilon_2$.\\
\indent It can be expected the match will remain to higher genus. Therefore, the 230 relations have already uniquely determined the elliptic genus of three E-strings. In other word, $Z_3$ does not contain any unfixed parameter even though $N_{\ydiagram{3}}$, $N_{\ydiagram{2,1}}$ and $N_{\ydiagram{1,1,1}}$ do contain. This is guaranteed by an identity among Jacobi forms, which will be shown later.\\
\indent The domain wall blocks with four unfixed parameters are quite complicated. In the massless case, we find that the four parameters can actually combine into a single parameter. By properly choosing the combination, we can make domain wall blocks only contain this one free parameter and exhibit some nice properties. We list our observations for massless case as follows:
\begin{enumerate}
\item $N_{\ydiagram{3}}(0)=N_{\ydiagram{1,1,1}}(0)=N_{\ydiagram{2,1}}(0,0)=-E_4^3$. This is quite similar with the situation of two E-strings where we have $N_{\ydiagram{2}}(0)=N_{\ydiagram{1,1}}(0)=E_4^2$. This is as expected, since all $\epsilon_i$ vanish means there is no twist at all.
\item The four parameters in domain wall blocks can combine into one single parameter $X$. See Appendix for one choice of $X$. If $X=0$, the domain wall block will get remarkable simplification. For example, the first order of the numerators with respect to $Q_{\tau}$ is just $-1$, the expansion coefficients only have denominators $2,3,6$. For generic $X$, the expression of the numerators will be more complicated. But the coefficients still only have denominators $2,3,6$. It can be shown impossible to make all coefficients integral.\\
    Under the condition $X=0$, the numerators of massless domain wall blocks have simple expressions:
   \begin{displaymath}
   \begin{split}
    &N_{\ydiagram{3}}^{\text{massless}}=\frac{1}{2^{18}3^{10}}\Big(-36 E_4^3 \phi_{0,1}^8+288 E_4^2 E_6 \phi_{0,1}^7 \phi_{-2,1}-E_4 \left(85 E_4^3+923 E_6^2\right) \phi_{0,1}^6 \phi_{-2,1}^2\\
    &\quad+96 E_6 \left(11 E_4^3+10 E_6^2\right) \phi_{0,1}^5 \phi_{-2,1}^3-9 E_4^2 \left(141 E_4^3+139 E_6^2\right) \phi_{0,1}^4 \phi_{-2,1}^4\\
    &\quad+8 E_4 E_6 \left(535 E_4^3-283 E_6^2\right) \phi_{0,1}^3 \phi_{-2,1}^5-3 \left(501 E_4^6+475 E_4^3 E_6^2-640 E_6^4\right) \phi_{0,1}^2 \phi_{-2,1}^6\\
    &\quad+72 E_4^2 E_6 \left(25 E_4^3-21 E_6^2\right) \phi_{0,1} \phi_{-2,1}^7+E_4 \left(-243 E_4^6+143 E_4^3 E_6^2+64 E_6^4\right) \phi_{-2,1}^8\Big),
    \end{split}
    \end{displaymath}
    \begin{displaymath}
    \begin{split}
    &N_{\ydiagram{2,1}}^{\text{massless}}(\epsilon_1,\epsilon_2)=\frac{1}{2^{16}3^{10}}\bigg[\Big(-9 E_4^3 \psi_{0}^3+27 E_4^2 E_6 \psi_{-2} \psi_{0}^2+E_4 (11 E_4^3-38 E_6^2) \psi_{-2}^2 \psi_{0}\\
    &+E_6 (20 E_6^2-11 E_4^3) \psi_{-2}^3\Big) \varphi_{0}^5+\Big(45 E_4^2 E_6 \psi_{0}^3-E_4 (76 E_4^3+59 E_6^2) \psi_{-2} \psi_{0}^2\\
    &+5 E_6 (23 E_4^3+4 E_6^2) \psi_{-2}^2 \psi_{0}+3 E_4^2 (2 E_4^3-17 E_6^2) \psi_{-2}^3\Big) \varphi_{-2} \varphi_{0}^4\\
    &-\Big((7 E_4^4+83 E_6^2 E_4) \psi_{0}^3+10 E_6 (10 E_6^2-37 E_4^3) \psi_{-2} \psi_{0}^2+3 E_4^2 (119 E_4^3-29 E_6^2) \psi_{-2}^2 \psi_{0}\\
    &+6 E_4 E_6 (14 E_6^2-29 E_4^3) \psi_{-2}^3\Big) \varphi_{-2}^2 \varphi_{0}^3+ \Big(30 E_6 (E_4^3+2 E_6^2) \psi_{0}^3+3 E_4^2 (29 E_6^2\\
    &-119 E_4^3) \psi_{-2} \psi_{0}^2+10 E_4 E_6 (73 E_4^3-46 E_6^2) \psi_{-2}^2 \psi_{0}+(-63 E_4^6-187 E_6^2 E_4^3\\
    &+160 E_6^4) \psi_{-2}^3\Big) \varphi_{-2}^3\varphi_{0}^2+ \Big((6 E_4^5-51 E_4^2 E_6^2) \psi_{0}^3+E_4 E_6 (313 E_4^3-178 E_6^2) \psi_{-2} \psi_{0}^2\\
    &+(-684 E_4^6+869 E_6^2 E_4^3-320 E_6^4) \psi_{-2}^2 \psi_{0}+45 E_4^2 E_6 (3 E_4^3-2 E_6^2) \psi_{-2}^3\Big) \varphi_{-2}^4\varphi_{0}\\
    &+ \Big((7 E_6 E_4^4+2 E_6^3 E_4) \psi_{0}^3+(99 E_4^6-286 E_6^2 E_4^3+160 E_6^4) \psi_{-2} \psi_{0}^2\\
    &+27 E_4^2 E_6 (3 E_4^3-2 E_6^2) \psi_{-2}^2 \psi_{0}-9 E_4 (81 E_4^6-160 E_6^2 E_4^3+80 E_6^4) \psi_{-2}^3\Big)\varphi_{-2}^5\bigg].
    \end{split}
    \end{displaymath}
    Here we omit the elliptic parameter in the formula of $N_{\ydiagram{3}}$ since there is only one of $\epsilon_i$. In the formula of $N_{\ydiagram{2,1}}$, we denote $\phi_{0,1}(\omega_1)$, $\phi_{-2,1}(\omega_1)$, $\phi_{0,1}(\omega_2)$, $\phi_{-2,1}(\omega_2)$ as $\varphi_0$, $\varphi_{-2}$, $\psi_0$, $\psi_{-2}$ respectively for short. Recall also $\omega_1=(\epsilon_1-\epsilon_2)/2$ and $\omega_2=(\epsilon_1+\epsilon_2)/2$.\\
    The expansion coefficients of domain wall blocks with respect to $q=e^{2\pi i \tau}$, $q_i=e^{2\pi i \epsilon_i}$ are supposed to have physical significance. In the case of two strings, they are all integers and interpreted as BPS degeneracy. This lead to the proposal that M9 domain wall formulas may be related to the computation of the open topological partition function on some Calabi-Yau threefold, just like M5 domain wall formulas \cite{Haghighat:2014pva}. However, in our computation here, the situation is more subtle since there are unfixed parameters in domain walls which traces to the lack of orthogonal bases for weak Jacobi form with two elliptic parameters. The variability of coefficients make their physical meaning vague. One may expect to obtain fine results by properly choosing the unfixed parameters. However, under the present paradigm, it seems unlikely to make all expansion coefficients integral. We show the first few orders of the numerators of the domain walls under the simplifying condition $X=0$, which are the most succinct expression we can have for the massless case:\\
    \begin{displaymath}
\begin{split}
&N_{\ydiagram{3}}^{\textrm{massless}}(q_1)=-1+ q\left(\frac{1}{2 q_1^4}+\frac{67}{6 q_1^3}-\frac{340}{3 q_1^2}-\frac{811}{6 q_1}-\frac{739}{3}-\frac{811 q_1}{6}-\frac{340 q_1^2}{3}+\frac{67 q_1^3}{6}+\frac{q_1^4}{2}\right)\\
&+q^2\bigg(-\frac{1}{2 q_1^7}-\frac{32}{3 q_1^6}+\frac{93}{q_1^5}-\frac{1700}{q_1^4}-\frac{9768}{q_1^3}-\frac{22592}{q_1^2}-\frac{70953}{2 q_1}-\frac{121112}{3}\\
&\quad-\frac{70953 q_1}{2}-22592 q_1^2-9768 q_1^3-1700 q_1^4+93 q_1^5-\frac{32 q_1^6}{3}-\frac{q_1^7}{2}\bigg)+O\left(q^3\right)
\end{split}
\end{displaymath}
\begin{displaymath}
\begin{split}
&N_{\ydiagram{2,1}}^{\textrm{massless}}(q_1,q_2)=-1+q \Bigg(\bigg(\frac{q_2}{2}+\frac{1}{2}\bigg)\frac{1}{q_1^2}+\bigg(\frac{q_2^2}{2}-\frac{64 q_2}{3}-\frac{683}{6}+\frac{32}{3 q_2}\bigg)\frac{1}{q_1}+\bigg(\frac{q_2^2}{2}\\
&\quad-\frac{683 q_2}{6}-\frac{742}{3}-\frac{683}{6 q_2}+\frac{1}{2 q_2^2}\bigg)+\left(\frac{32 q_2}{3}-\frac{683}{6}-\frac{64}{3 q_2}+\frac{1}{2 q_2^2}\right) q_1+\left(\frac{1}{2}+\frac{1}{2 q_2}\right) q_1^2\Bigg)\\
&+q^2 \Bigg(-\frac{q_2}{q_1^4}+\bigg(\frac{32 q_2^2}{3}-\frac{683 q_2}{6}-\frac{64}{3}+\frac{1}{2 q_2}\bigg)\frac{1}{q_1^3}\\
&\quad+\bigg(\frac{32 q_2^3}{3}+\frac{619 q_2^2}{3}-1917 q_2-1917+\frac{619}{3 q_2}+\frac{32}{3 q_2^2}\bigg)\frac{1}{q_1^2}\\
&\quad+\bigg(-q_2^4-\frac{683 q_2^3}{6}-1917 q_2^2-14493 q_2-\frac{62641}{3}-\frac{23489}{3 q_2}+\frac{619}{3 q_2^2}+\frac{1}{2 q_2^3}\bigg)\frac{1}{q_1}\\
&\quad+\left(-\frac{64 q_2^3}{3}-1917 q_2^2-\frac{62641 q_2}{3}-\frac{109610}{3}-\frac{62641}{3 q_2}-\frac{1917}{q_2^2}-\frac{64}{3 q_2^3}\right)\\
&\quad+\left(\frac{q_2^3}{2}+\frac{619 q_2^2}{3}-\frac{23489 q_2}{3}-\frac{62641}{3}-\frac{14493}{q_2}-\frac{1917}{q_2^2}-\frac{683}{6 q_2^3}-\frac{1}{q_2^4}\right)q_1\\
&\quad+\left(\frac{32 q_2^2}{3}+\frac{619 q_2}{3}-1917-\frac{1917}{q_2}+\frac{619}{3 q_2^2}+\frac{32}{3 q_2^3}\right) q_1^2\\
&\quad+\left(\frac{q_2}{2}-\frac{64}{3}-\frac{683}{6 q_2}+\frac{32}{3 q_2^2}\right) q_1^3-\frac{q_1^4}{q_2}\Bigg)+O\left(q^4\right)
\end{split}
\end{displaymath}
Note $q_1$ and $q_2$ is actually symmetric in $N_{\ydiagram{2,1}}(q_1,q_2)$, as expected.
\item The numerator $N_{\ydiagram{2,1}}$ has integral and fixed expansion automatically in Nekrasov-Shatashvili limit $\epsilon_2=0$. By automatically, we mean the unfixed parameters in $N_{\ydiagram{2,1}}$ cancel out in the limit. This phenomenon is shared by the massive case which can be checked directly. The physical meaning of this is not clear so far.\\
    Taking $q_2=1$, we show the first few orders of massless $N_{\ydiagram{2,1}}$ in NS limit:\\
    \begin{displaymath}
\begin{split}
&N_{\ydiagram{2,1}}^{\mathrm{massless}}(q_1,1)=-1+q \left(q_1^2+\frac{1}{q_1^2}-124 q_1-\frac{124}{q_1}-474\right)\\
&+q^2 \left(-q_1^4-\frac{1}{q_1^4}-124 q_1^3-\frac{124}{q_1^3}-3400 q_1^2-\frac{3400}{q_1^2}-45028 q_1-\frac{45028}{q_1}-82174\right)\\
&+q^3 \bigg(-474 q_1^4-\frac{474}{q_1^4}-45028 q_1^3-\frac{45028}{q_1^3}-970980 q_1^2-\frac{970980}{q_1^2}\\
&-4180524 q_1-\frac{4180524}{q_1}-6560548\bigg)+O\left(q^4\right)
\end{split}
\end{displaymath}
\end{enumerate}
\indent One possible reason for this non-integral situation may be that we are not working on perfect bases. Apparently, the 170 bases are not completely independent for weak Jacobi form with index quadratic form $2\epsilon_1^2-\epsilon_1\epsilon_2+2\epsilon_2^2$. We expect the perfect bases for weak Jacobi form with two elliptic parameters must contain some bases with two elliptic parameters. Of course, their explicit formulas are still unknown. Therefore, we assume here the index matrix can be transformed into a diagonal one and then we only need the bases with one parameter. Intuitively, this procedure will increase the number of bases. That is why we only find 230 relations among 234 coefficients. The consequence is that we can guarantee the results which involve the combination between $N_{\ydiagram{3}}$ and $N_{\ydiagram{2,1}}$ right and fixed, such as the elliptic genus of three E-strings and three H-strings, but each domain wall block can change and does not have integral expansion.\\
\indent Another possibility is that even if we have perfect bases for $N_{\ydiagram{2,1}}$, the numerators still will not have integral expansion and they probably have the same form with the present results with condition $X=0$ in massless case. There may be some physical reasons for the denominators. Besides, the whole domain walls may show some fine properties that the numerators do not, since there are still many other ingredients in the expression of domain walls, see Eq \eqref{eq:M9ansatz}.\\
\indent Now we make some remarks on the massive case. Since there are only 230 relations among 234 parameters, we can choose four parameters as variables and express other parameters as polynomials of them. We find that in the domain wall blocks $N_{\ydiagram{3}}$ and $N_{\ydiagram{2,1}}$, the terms proportional to $A_1^3$ are already fixed by the 230 relations, and if we choose $c_{212}$, $c_{39}$, $c_{415}$ and $c_{514}$ as free variables, then the terms proportional to $A_1A_2$, $A_1B_2$, $A_3$ and $B_3$ only contain $c_{212}$, $c_{39}$, $c_{415}$ and $c_{514}$ respectively. This choice is quite natural if we obtain the relations in the match of free energy order by order.\\
\indent Since our results show the elliptic genus \eqref{eq:3Eellipticgenus} actually contains no unfixed parameters, this indicates if we extract the terms proportional to $A_1A_2$, $A_1B_2$, $A_3$ and $B_3$ respectively in \eqref{eq:3Eellipticgenus}, we will obtain four equations and each of them contains one free parameter. The terms with free parameter as coefficient in each equation are supposed to vanish when we sum over all level-three Young diagrams. To assure our formula for the elliptic genus is indeed definite, this should be proved. In fact, we find all vanishments actually are guaranteed by one identity among relevant Jacobi forms, which is also the reason why the four unfixed parameters can combine into one single parameter. Denote
\begin{displaymath}
\begin{split}
M_1=&2E_4 \phi_{-2,1} \Big[\phi_{0,1}^7 - 9 E_4 \phi_{0,1}^5 \phi_{-2,1}^2 +
   10 E_6 \phi_{0,1}^4 \phi_{-2,1}^3+ 15 E_4^2 \phi_{0,1}^3 \phi_{-2,1}^4 -
   36 E_4 E_6 \phi_{0,1}^2 \phi_{-2,1}^5\\
    &+ (9 E_4^3 + 16 E_6^2) \phi_{0,1} \phi_{-2,1}^6- 6 E_4^2 E_6 \phi_{-2,1}^7\Big],
\end{split}
\end{displaymath}
\begin{displaymath}
\begin{split}
M_2=&E_4 (\psi_0 \varphi_2 -
   \varphi_0 \psi_2) \Big[8 \varphi_0^3 \varphi_2 \psi_2 (E_4 \psi_0 - E_6 \psi_2)
+    \varphi_0^4 (-\psi_0^2 + E_4 \psi_2^2)\\
& + 6 \varphi_0^2 \varphi_2^2 (E_4 \psi_0^2 -
      4 E_6 \psi_0 \psi_2 +
      3 E_4^2 \psi_2^2) -
   8 \varphi_0 \varphi_2^3 (E_6 \psi_0^2 -
      3 E_4^2 \psi_0 \psi_2 +
      2 E_4 E_6 \psi_2^2)\\
      & +
   \varphi_2^4 (3 E_4^2 \psi_0^2 -
      8 E_4 E_6 \psi_0 \psi_2 + (-27 E_4^3 + 32 E_6^2) \psi_2^2)\Big].
\end{split}
\end{displaymath}
Then the identity can be written as:
\begin{equation}\label{eq:identity}
\begin{split}
&\frac{M_1(\epsilon_1)}{\theta_1(3\epsilon_1)\theta_1(2\epsilon_1-\epsilon_2)
\theta_1(2\epsilon_1)\theta_1(\epsilon_1-\epsilon_2)}+\frac{M_1(\epsilon_2)}{\theta_1(\epsilon_1-2\epsilon_2)\theta_1(-3\epsilon_2)
\theta_1(\epsilon_1-\epsilon_2)\theta_1(-2\epsilon_2)}\\
&\quad\quad\quad\quad\quad\quad\quad+\frac{M_2(\epsilon_1,\epsilon_2)}{\theta_1(2\epsilon_1-\epsilon_2)
\theta_1(\epsilon_1-2\epsilon_1)\theta_1(\epsilon_1)\theta_1(-\epsilon_2)}=0 . 
\end{split}
\end{equation}
It should be possible to directly prove the identity using the addition formulas of Jacobi theta functions. Here we will not give the detailed proof since it is already confirmed by extensive checks on series expansion. These checks are highly nontrivial and include this identity as special case. It is also easy to check Eq \eqref{eq:identity} numerically. \\
\indent In the appendix, we list the 234 coefficients which are solved from 230 relations from the match of free energy. These are enough to uniquely determine the elliptic genus of three E-strings and three H-strings.\\
\subsection{Elliptic genus of three H-strings}
The M9 domain walls can also be used to compute the three heterotic string partition function. In the Ho\v rava-Witten picture of $E_8\times E_8$ heterotic string theory, H-strings are realized by M2 branes stretched between two M9 branes. Therefore, we expect the H-string amplitude can be written as some combination of left M9 domain walls and right M9 domain walls. Note that
\begin{equation}
\begin{split}
&D_{\ydiagram{3}}^{M9,L}(\vec{m}_L)D_{\ydiagram{3}}^{M9,R}(\vec{m}_R)\\
=&-\frac{N_{\ydiagram{3}}^{M9,L}(\vec{m}_L,\epsilon_1)
N_{\ydiagram{3}}^{M9,R}(\vec{m}_R,\epsilon_1)}{\eta^{36}\theta_1(3\epsilon_1)\theta_1(2\epsilon_1-\epsilon_2)
\theta_1(2\epsilon_1)\theta_1(\epsilon_1-\epsilon_2)\theta_1(\epsilon_1)\theta_1(\epsilon_2)}\cdot\\
&\frac{1}{\theta_1(\epsilon_3-2\epsilon_1)\theta_1(\epsilon_3-\epsilon_1)
\theta_1(\epsilon_3)\theta_1(\epsilon_4)\theta_1(\epsilon_1-\epsilon_4)\theta_1(2\epsilon_1-\epsilon_4)},
\end{split}
\end{equation}
and
\begin{equation}
\begin{split}
&D_{\ydiagram{2,1}}^{M9,L}(\vec{m}_L)D_{\ydiagram{2,1}}^{M9,R}(\vec{m}_R)\\
=&-\frac{N_{\ydiagram{2,1}}^{M9,L}(\vec{m}_L,\epsilon_1)
N_{\ydiagram{2,1}}^{M9,R}(\vec{m}_R,\epsilon_1)}{\eta^{36}
\theta_1(2\epsilon_1-\epsilon_2)\theta_1(\epsilon_1-2\epsilon_2)
\theta_1^2(\epsilon_1)\theta_1^2(\epsilon_2)}\cdot\\
&\frac{1}{\theta_1(\epsilon_3-\epsilon_1)\theta_1(\epsilon_3)
\theta_1(\epsilon_3-\epsilon_2)\theta_1(\epsilon_1-\epsilon_4)
\theta_1(\epsilon_4)\theta_1(\epsilon_2-\epsilon_4)}.
\end{split}
\end{equation}
Considering the symmetry among $\epsilon_{1,2,3,4}$ for H-strings, we suggest the following formula for the elliptic genus for three heterotic strings:
\begin{equation}\label{eq:3Hdomainwall}
\begin{split}
Z_3^{\mathrm{Het}}=&\Big(D_{\ydiagram{3}}^{M9,L}(\vec{m}_L)D_{\ydiagram{3}}^{M9,R}(\vec{m}_R)
+(\epsilon_1\leftrightarrow\epsilon_2)+(\epsilon_1\leftrightarrow\epsilon_3)
+(\epsilon_1\leftrightarrow\epsilon_4)\Big)\\
&\Big(D_{\ydiagram{2,1}}^{M9,L}(\vec{m}_L)D_{\ydiagram{2,1}}^{M9,R}(\vec{m}_R)
+(\epsilon_1,\epsilon_2\leftrightarrow\epsilon_1,\epsilon_3)
+(\epsilon_1,\epsilon_2\leftrightarrow\epsilon_1,\epsilon_4)\\
&+(\epsilon_1,\epsilon_2\leftrightarrow\epsilon_2,\epsilon_3)
+(\epsilon_1,\epsilon_2\leftrightarrow\epsilon_2,\epsilon_4)
+(\epsilon_1,\epsilon_2\leftrightarrow\epsilon_3,\epsilon_4)\Big).
\end{split}
\end{equation}
\indent On the other hand, it is well-known the elliptic genus of $n$ heterotic strings can be computed from Hecke transformation, as we have reviewed in Section 3.1. In the case of three strings,
\begin{equation}\label{eq:3Horbifold}
\begin{split}
Z_3^{\mathrm{Het}}=&\frac{1}{3}\bigg(Z_1^{\mathrm{Het}}(3\tau,3\epsilon_i,3\vec{m})
+Z_1^{\mathrm{Het}}(\frac{\tau}{3},\epsilon_i,\vec{m})+Z_1^{\mathrm{Het}}(\frac{\tau+1}{3},\epsilon_i,\vec{m})
+Z_1^{\mathrm{Het}}(\frac{\tau+2}{3},\epsilon_i,\vec{m})\bigg)\\
&+\frac{1}{2}Z_1^{\mathrm{Het}}(\tau,\epsilon_i,\vec{m})\Big(Z_1^{\mathrm{Het}}(2\tau,2\epsilon_i,2\vec{m})
+Z_1^{\mathrm{Het}}(\frac{\tau}{2},\epsilon_i,\vec{m})+Z_1^{\mathrm{Het}}(\frac{\tau+1}{2},\epsilon_i,\vec{m})\Big)\\
&+\frac{1}{6}Z_1^{\mathrm{Het}}(\tau,\epsilon_i,\vec{m})^3.
\end{split}
\end{equation}
\indent We checked $3E+3E\rightarrow3H$ exactly for the first order $Q_{\tau}^{-3}$, numerically up to order $Q_{\tau}^{1}$ for massless case and up to order $Q_{\tau}^{0}$ for massive case. We make some observations as the followings
\begin{enumerate}
\item For the first order ($Q_{\tau}^{-3}$) of $3E+3E=3H$, which does not contain any $m_i$, both sides do not change as the four remaining free parameters vary. The first order identity holds even without the constrain $\epsilon_1+\epsilon_2+\epsilon_3+\epsilon_4=0$.
\item The higher orders of $3E+3E=3H$ only hold under the condition $\epsilon_1+\epsilon_2+\epsilon_3+\epsilon_4=0$. Both sides also do not change as the four remaining free parameters vary.
\end{enumerate}
Since the M9 domain walls depend linearly on four free parameters, the left side of $3E+3E=3H$ is a quadratic polynomial of the four free parameters on the surface. Except for the constant term, all the coefficients of the polynomial are supposed to vanish when we sum up in Eq \eqref{eq:3Hdomainwall}. This will lead to some identities like Eq \eqref{eq:identity}. To save space, we do not  show the explicit expression of these identities here, but regard them as implications of our main identity $3E+3E=3H$.\\
\indent Note each term in Eq \eqref{eq:3Hdomainwall} exhibits distinct and non-diagonal index quadratic form with respect to $\epsilon_i$ on the surface. However, all index quadratic forms of these terms are actually equivalent to the simple diagonal index quadratic form in Eq \eqref{eq:3Horbifold} under our symmetry preserving condition $\sum_i\epsilon_i=0$. For example, let us consider the first term in Eq \eqref{eq:3Hdomainwall}. Since the indices for $m_i$ obviously match for the two sides of $3E+3E=3H$, we only show the index quadratic form for $\epsilon_i$. Denote $\partial_{E_2}^{\epsilon}$ as the operator for the modular anomaly relevant to $\epsilon_i$, then\\
\begin{equation}
\begin{split}
&\partial_{E_2}^{\epsilon}\log \left(D_{\ydiagram{3}}^{M9,L}(\vec{m}_L)D_{\ydiagram{3}}^{M9,R}(\vec{m}_R)\right)\\
=&\ 2\partial_{E_2}^{\epsilon}\log N_{\ydiagram{3}}^{M9}(\epsilon_1)
-\partial_{E_2}^{\epsilon}\log
\Big(\theta_1(3\epsilon_1)\theta_1(2\epsilon_1-\epsilon_2)
\theta_1(2\epsilon_1)\theta_1(\epsilon_1-\epsilon_2)\theta_1(\epsilon_1)\theta_1(\epsilon_2)\Big)\\
&-\partial_{E_2}^{\epsilon}\log\Big(\theta_1(\epsilon_3-2\epsilon_1)\theta_1(\epsilon_3-\epsilon_1)
\theta_1(\epsilon_3)\theta_1(\epsilon_4)\theta_1(\epsilon_1-\epsilon_4)\theta_1(2\epsilon_1-\epsilon_4)\Big)\\
=&-\frac{1}{24}\Big(32\epsilon_1^2-\big((3\epsilon_1)^2+(2\epsilon_1-\epsilon_2)^2+
(2\epsilon_1)^2+(\epsilon_1-\epsilon_2)^2+\epsilon_1^2+\epsilon_2^2\\
&+(\epsilon_3-2\epsilon_1)^2+(\epsilon_3-\epsilon_1)^2+
\epsilon_3^2+\epsilon_4^2+(\epsilon_1-\epsilon_4)^2+(2\epsilon_1-\epsilon_4)^2\big)\Big)\\
=&-\frac{3}{24}\Big(\epsilon_1^2-\epsilon_2^2-\epsilon_3^2-\epsilon_4^2+2\epsilon_1\epsilon_2+2\epsilon_1\epsilon_3+2\epsilon_1\epsilon_4\Big).
\end{split}
\end{equation}
The modular anomaly of the other terms in Eq \eqref{eq:3Hdomainwall} can be computed similarly. On the other hand, the property of Hecke transformation guarantees all indices of $Z_3^{\mathrm{Het}}$ are just three times as those of $Z_1^{\mathrm{Het}}$. Thus,
\begin{equation}
\begin{split}
&\partial_{E_2}^{\epsilon}\log Z_3^{\mathrm{Het}}=-\frac{3}{24}\Big(\epsilon_1^2+\epsilon_2^2+\epsilon_3^2+\epsilon_4^2\Big).
\end{split}
\end{equation}
These two results are obviously equivalent if we apply the condition $\sum_i\epsilon_i=0s$. See a more general statement for the relation between the modular anomaly for E-strings, M-strings and H-strings in \cite{Haghighat:2014pva}.\\
\indent Now we present the first order identity of $3E+3E\rightarrow 3H$ to show how the two formulations are different on the surface. Denote $q_i=e^{2\pi i\epsilon_i}$, then from Eq \eqref{eq:3Horbifold} it is easy to show the first order of elliptic genus of three heterotic strings computed from orbifold formula is
\begin{equation}\label{eq:3Horbifoldfirstorder}
\begin{split}
Z_3^{\mathrm{H(1st)}}=-\frac{1}{6}\left[\frac{1}{\prod_{k=1}^4(1-q_k)^3}
-3\frac{1}{\prod_{k=1}^4(1-q_k^2)(1-q_k)}+2\frac{1}{\prod_{k=1}^4(1-q_k^3)}\right].
\end{split}
\end{equation}
From Eq \eqref{eq:3Hdomainwall} we obtain the first order of elliptic genus of three heterotic strings computed from domain wall method:
\begin{equation}\label{eq:3Hdomainwallfirstorder}
\begin{split}
Z_3^{\mathrm{H(1st)}}=&
-\Bigg[\frac{q_1^6}{\big(\prod_{k=1}^4(1-q_k)\big)(1-q_1^2)(1-q_1^3)\big(\prod_{k=2}^4(q_1-q_k)(q_1^2-q_k)\big)}
+(\textrm{3 permutations})\Bigg]\\
&-\Bigg[\frac{q_1^2q_2^2}{\big(\prod_{k=1}^4(1-q_k)\big)
(1-q_1)(1-q_2)(q_1-q_2^2)(q_2-q_1^2)\big(\prod_{k=3}^4(q_1-q_k)(q_2-q_k)\big)}\\
&\quad+(\textrm{5 permutations})\Bigg].
\end{split}
\end{equation}
The above two expressions can be shown identical by direct check. This is already conjectured by the authors of \cite{Haghighat:2014pva}.\\
\indent Note also Eq \eqref{eq:3Horbifoldfirstorder} and \eqref{eq:3Hdomainwallfirstorder} are identical with arbitrary $q_i$. This is quite similar with the situation of two strings, where the first order of $2E+2E\rightarrow 2H$ holds even without the constraint $q_1q_2q_3q_4=1$.\\
\indent We make some further comments on the two expressions for the elliptic genus of three genus. The expression from orbifold formula has simple pole structure and explicit index matrix for $\epsilon_i$. The indices for $\epsilon_i$ and $\vec{m}_{E_8\times E_8}$ are guaranteed by the property of Hecke transformation. Each term in Eq \eqref{eq:3Horbifold} may not have the index matrix as requested. Besides, the left and right $E_8$ masses are entangled in a nontrivial way.\\
\indent On the other hand, the expression constructed from M9 domain walls has more complicated pole structure. However, it is easy to recognize that the irregular poles actually will cancel out among the permutations and only the poles same as those in orbifold formula remain. Each term in this expression has different index matrix for $\epsilon_i$ on the surface. But once using the constraint $\epsilon_1+\epsilon_2+\epsilon_3+\epsilon_4=0$, they will result in the same and right index matrix which is diagonal. Besides, in this formula, one can perform independent modular transformation on the left and right degree of freedom.
\subsection{Discussion of results}
So far we have derived the M9 domain walls of level three and use the domain wall blocks to construct the elliptic genera. This leads for the first time to a precise expression for the elliptic genus of three E-strings and a new formula the elliptic genus of three heterotic strings. The equivalence between the two expressions of elliptic genus of heterotic strings, or $3E+3E\rightarrow3H$ for short, involves highly nontrivial identities among Jacobi forms.\\
\indent M5 domain walls are known to be equal to the open topological string partition function for a certain toric Calabi-Yau threefold. Therefore, it is natural to ask whether M9 domain walls correspond to some open topological string partition function for some Calabi-Yau threefold. If this is correct, we expect the domain walls will have integral expansions since the coefficients can be interpreted as BPS degeneracy. In the case of two strings, the domain walls do have integral expansion. However, in the case of three strings, as we have show in Section 3.3, the domain walls (at least the numerators) do not have integral expansions. The best we can do is making the expansion coefficients only have denominators $2,3,6$. Since we have checked $3E+3E\rightarrow 3H$ right, there must be some reasons for our coefficients to be right. If we insist on the BPS degeneracy interpretation, then one possible reason might be the symmetrization in Eq \eqref{eq:3Hdomainwall} contain some weights. We will study further to determine whether there is integral expansion in other conventions.\\
\indent We also wish to make some remarks about more strings. In the case of more than three strings, the main problem is still the non-diagonal index matrix. For example, in the case of four strings, it is easy to calculated the following modular anomaly for the numerators of domain walls blocks of level four partition:
\begin{equation}
\partial_{E_2}\log N_{\ydiagram{4}}(\epsilon_1,\epsilon_2)=-\frac{1}{24}\Big(40\epsilon_1^2+4\sum m_i^2\Big),
\end{equation}
\begin{equation}
\partial_{E_2}\log N_{\ydiagram{3,1}}(\epsilon_1,\epsilon_2)=
-\frac{1}{24}\Big(16\epsilon_1^2-4\epsilon_1\epsilon_2+4\epsilon_2^2+4\sum m_i^2\Big),
\end{equation}
\begin{equation}
\partial_{E_2}\log N_{\ydiagram{2,2}}(\epsilon_1,\epsilon_2)=-\frac{1}{24}\Big(8\epsilon_1^2+8\epsilon_2^2+3\sum m_i^2\Big).
\end{equation}
Obviously, the key point lies in $ N_{\ydiagram{3,1}}$. We tried to transform its index quadratic form as $5(\epsilon_1-\epsilon_2/2)^2+3(\epsilon_1+\epsilon_2/2)^2$, but this ansatz turns out to be not right, as we can not match the resulting expression of elliptic genus with free energy to high orders. So we cannot expect this method to work all the time.\\
\indent We also noticed the first order of $4E+4E\rightarrow 4H$ does not holds if we assume $N_{\ydiagram{4}}\sim N_{\ydiagram{3,1}}\sim N_{\ydiagram{2,2}}\sim 1+O(Q_{\tau})$.\footnote{This is also noticed by the authors of \cite{Haghighat:2014pva}.} One possibility is that the first order of these numerators are not just $1$ but some functions of $\epsilon_i$. The other reason may be that when we do the computations for E-strings, we miss part of the information of M9 domain walls since we need to project them to the state of M5 branes.\\
\indent Of course, many subtle issues here are due to the lack of orthogonal bases for weak Jacobi forms with generic index quadratic form. If we do find the perfect bases, we can easily perform the calculation to arbitrary $n$ strings.

\acknowledgments

We thank Guglielmo Lockhart, Jae-Hyun Yang and especially Babak Haghighat for useful discussions and correspondences. We thank Albrecht Klemm for reading the draft. MH thanks the organizers of a workshop on "String Theory and its Applications"  in IHEP, Beijing for hospitality, and  Kimyeong Lee for discussions at the workshop. While we are preparing the draft, the paper \cite{Kim:2014} appears, whose results overlap with ours but the methods are different. MH is supported by the ``Young Thousand People" plan by the Central Organization Department in China, and by the Natural Science Foundation of China.

\appendix
\section{Coefficient List}
The following values of the coefficients in the domain wall blocks are solved from 230 relations obtained from the match of free energy. Some of them can be determined to numbers, the others can be expressed by four parameters $c_{212}$, $c_{39}$, $c_{415}$ and $c_{514}$, which are denoted as $x,y,z,w$ respectively for short. Of course, we can choose other four parameters as free variables. It does not matter. These values or relations are enough to uniquely determine elliptic genus. For example, we can simply choose $x=y=z=w=0$ to obtain the exact elliptic genus. We keep them here because they may indicate some interesting issues when we do have the orthogonal bases for Jacobi forms of higher degrees. For the massless case, the domain wall blocks simplify dramatically under the constraint $X=-5003/15552 + 995328(x-3y/8+z+w)=0$. We have shown the explicit expression of domain walls under this simplifying condition. For the massive case, this condition brings no significant simplification since there are still other three parameters in the domain walls. The expressions in the Appendix do not impose this simplifying condition.
\\
\indent We also notice an interesting fact that the denominators of all these coefficients only contain factors 2 and 3. This phenomenon is shared by the domain walls for the case of two strings.\\
\newpage
\begin{displaymath}
\begin{split}
&a_{111} = -1/429981696,\ c_{11} = -1/429981696,\  a_{113} = 5/429981696,\ a_{122} = -7/429981696,\\
&a_{131} = 1/71663616,\ a_{114} = -1/107495424,\  a_{123} = 1/107495424,\ a_{132} = 5/53747712,\\
&a_{141} = -1/53747712,\  a_{124} = 1/143327232,\ a_{133} = -13/71663616,\ a_{142} = -13/71663616,\\
&a_{151} = 1/143327232,\  c_{12} = 1/23887872,\  a_{211} = 0,\ c_{21} = 0,\\
&a_{214} = 11/1719926784 + x/12,\ a_{212} = 1/1719926784 - x/12,\\
&a_{221} = 11/1719926784 + x/12,\ a_{312} = 125/6879707136 + y/32,\\
&a_{321} = -5/764411904 - y/32,\ a_{313} = -5/429981696,\\
&a_{322} = -5/429981696,\ a_{331} = -5/214990848,\\
&a_{314} = -5/764411904 - y/32,\ a_{323} = -185/2293235712 - 7 y/32,\\
&a_{332} = 5/382205952 + y/16,\ a_{341} = 125/1146617856 + 3 y/16,\\
&a_{324} = 85/859963392 + y/4,\ a_{333} = 125/429981696 + y/2,\\
&a_{342} = -65/429981696 - y/2,\ a_{351} = -35/286654464 - y/4,\\
&a_{412} = 7/1451188224 - z/12,\ a_{421} = -7/1451188224 + z/12,\\
&a_{512} = 5/483729408 - w/12,\ a_{521} = -5/483729408 + w/12,\\
&a_{514} = -5/322486272 + w/12,\ a_{523} = -25/322486272 + 7 w/12,\\
&a_{532} = 5/107495424 - w/6,\ a_{541} = 5/107495424 - w/2,\\
&b_{224} = -5/214990848 - 2 x/3,\ b_{233} = -1/13436928 - 4 x/3,\\
&b_{242} = -1/107495424 + 4 x/3,\ b_{251} = 1/107495424 + 2 x/3,\\
&b_{413} = 7/3869835264,\ b_{422} = -7/1934917632,\ b_{431} = 7/3869835264,\\
&b_{424} = 511/11609505792 - 2 z/3,\ b_{433} = 833/11609505792 - 4 z/3,\\
&b_{442} = -959/11609505792 + 4 z/3,\ b_{451} = -385/11609505792 + 2 z/3,\\
&a_{223} = 41/1719926784 + 7 x/12,\  a_{232} = -11/859963392 - x/6,\\
&a_{241} = 1/286654464 - x/2,\  a_{213} = -1/143327232,\ a_{222} = -1/143327232,\\
&a_{231} = -1/71663616,\  a_{224} = 0,\ a_{233} = 1/11943936,\  a_{242} = 1/11943936,\\
&a_{251} = 0,\ c_{22} = 5/1719926784 - x/6,\  c_{23} = -23/573308928,\\
&c_{31} = 5/127401984 + y/16,\  a_{411} = 0,\  c_{41} = 0,\ a_{413} = -7/3869835264,\\
&a_{422} = 7/1934917632,\ a_{431} = -7/3869835264,\  a_{414} = -7/1451188224 + z/12,\\
&a_{423} = -49/1451188224 + 7 z/12,\  a_{432} = 7/725594112 - z/6,\\
&a_{441} = 7/241864704 - z/2,\  a_{424} = -7/1289945088,\\
&a_{433} = 7/1289945088,\  a_{442} = 7/1289945088,\ a_{451} = -7/1289945088,\\
&c_{42} = 7/725594112 - z/6,\ c_{43} = -7/967458816,\  a_{511} = 0,\  c_{51} = 0,\\
&a_{513} = 0,\ a_{522} = 0,\  a_{531} = 0,\  b_{514} = 5/967458816,\\
&b_{523} = 5/967458816,\  b_{532} = -25/967458816,\ b_{541} = 5/322486272,\\
&a_{524} = 5/60466176 - 2 w/3,\ a_{533} = 5/30233088 - 4 w/3,\  a_{542} = -5/30233088 + 4 w/3,\\
\end{split}
\end{displaymath}
\begin{displaymath}
\begin{split}
&a_{551} = -5/60466176 + 2 w/3,\  c_{52} = 5/241864704 - w/6,\\
&a_{134} = 1/13436928,\  a_{143} = 5/13436928,\  a_{152} = 1/6718464,\ a_{161} = 0,\\
&a_{234} = 5/286654464 + 3 x/2,\ a_{243} = -29/286654464 + x/2,\\
&a_{252} = -49/573308928 - 7 x/4,\ a_{261} = -1/191102976 - x/4,\\
&a_{334} = -125/382205952 - 9 y/16,\ a_{343} = -5/127401984 - 3 y/16,\\
&a_{352} = 185/764411904 + 21 y/32,\ a_{361} = 5/254803968 + 3 y/32,\\
&a_{434} = -7/322486272,\ a_{443} = 35/967458816,\ a_{452} = -7/967458816,\\
&a_{461} = -7/967458816,\ a_{534} = -5/35831808 + 3 w/2,\ a_{543} = -5/35831808 + w/2,\\
&a_{552} = 25/107495424 - 7 w/4,\ a_{561} = 5/107495424 - w/4,\  b_{334} = 5/107495424,\\
&b_{343} = -25/107495424,\  b_{352} = -5/214990848,\ b_{361} = 5/214990848,\\
&b_{434} = -7/107495424 + 3 z/2,\ b_{443} = -7/107495424 + z/2,\\
&b_{452} = 35/322486272 - 7 z/4,\ b_{461} = 7/322486272 - z/4,\\
&b_{534} = -5/107495424,\ b_{543} = 25/322486272,\  b_{552} = -5/322486272,\\
&b_{561} = -5/322486272,\  c_{13} = -7/107495424,\ c_{14} = -5/35831808,\\
&c_{24} = 7/63700992 + 3 x/2,\ c_{25} = 11/191102976,\  c_{32} = -115/1719926784,\\
&c_{33} = -145/1146617856 - 9 y/16,\  c_{34} = 25/95551488 + 5 y/8,\\
&c_{36} = 25/214990848,\  c_{44} = 7/967458816,\  c_{48} = -7/483729408,\\
&c_{45} = -7/80621568 + 3 z/2,\  c_{46} = 0,\  c_{53} = 0,\ c_{54} = -5/20155392 + 3 w/2,\\
&c_{56} = 25/120932352 - 5 w/3,\ c_{55} = 5/80621568,\  a_{144} = 1/7962624,\\
&a_{153} = -7/47775744,\ a_{162} = 5/47775744,\  a_{244} = -1/7962624,\  a_{253} = -1/15925248,\\
&a_{262} = -1/15925248,\  a_{344} = 5/15925248 + y/2,\ a_{353} = 5/95551488 - y/4,\\
&a_{362} = -25/286654464 - y/4,\ a_{444} = 1085/11609505792 - 4 z/3,\\
&a_{453} = -413/5804752896 + 2 z/3,\ a_{462} = -259/11609505792 + 2 z/3,\\
&a_{544} = 5/120932352,\ a_{553} = -5/60466176,\  a_{562} = 5/120932352,\\
&b_{144} = -1/3359232,\  b_{153} = -1/6718464,\  b_{162} = -1/6718464,\\
&b_{244} = 1/6718464 - 4 x/3,\  b_{253} = 19/107495424 + 2 x/3,\\
&b_{262} = 5/53747712 + 2 x/3,\  b_{444} = -7/429981696,\ b_{453} = 7/214990848,\\
&b_{462} = -7/429981696,\ b_{544} = 5/40310784 - 4 w/3,\  b_{553} = 2 w/3,\\
&b_{562} = -5/40310784 + 2 w/3,\  c_{15} = 1/1990656,\ c_{16} = -5/23887872,\\
&c_{17} = -5/13436928,\ c_{26} = -23/214990848 - 5 x/3,\  c_{27} = -115/573308928 - 5 x/2,\\
&c_{28} = 1/21233664,\  c_{35} = 25/382205952 + 15 y/16,\ c_{410} = -133/322486272 + 6 z,\\
&c_{47} = 35/362797056 - 5 z/3,\ c_{411} = 7/107495424,\  c_{510} = -35/40310784 + 6 w,\\
&c_{58} = -5/20155392,\  c_{59} = 5/40310784,\  a_{154} = 5/35831808,\\
&a_{163} = 1/11943936,\  a_{254} = 7/191102976 - 9 x/4,\ a_{263} = 7/63700992 + 9 x/4,\\
&a_{354} = 5/28311552 + 27 y/32,\ a_{363} = -125/254803968 - 27 y/32,\\
\end{split}
\end{displaymath}
\begin{displaymath}
\begin{split}
&a_{454} = -7/45349632 + 8 z/3,\ a_{463} = 7/45349632 - 8 z/3,\\
&a_{554} = -5/15116544 + 8 w/3,\ a_{563} = 5/15116544 - 8 w/3,\\
&b_{254} = -1/13436928 + 8 x/3,\ b_{263} = -1/6718464 - 8 x/3,\\
&b_{354} = -35/107495424 - y,\ b_{363} = 55/107495424 + y,\  b_{454} = 7/53747712 - 9 z/4,\\
&b_{463} = -7/53747712 + 9 z/4,\  b_{554} = 5/17915904 - 9 w/4,\\
&b_{563} = -5/17915904 + 9 w/4,\  c_{18} = 11/35831808,\ c_{211} = -11/63700992 - 3 x/2,\\
&c_{29} = 7/23887872 + 6 x,\ c_{37} = -385/573308928 - 9 y/4,\\
&c_{38} = 65/127401984 + 9 y/16,\ c_{413} = -7/161243136 - 3 z/2,\\
&c_{49} = 77/483729408 - 5 z/2,\ c_{511} = 125/241864704 - 8 w/3,\\
&c_{57} = 5/8957952 - 5 w/2,\ a_{164} = -1/5308416,\  a_{264} = 0,\  a_{364} = 5/23887872,\\
&a_{464} = 0,\  a_{564} = 0,\  b_{164} = 1/6718464,\ b_{264} = 1/71663616,\\
&b_{364} = -5/26873856,\  b_{464} = 0,\ b_{564} = 0,\  c_{10} = 0,\  c_{19} = -1/15925248,\\
&c_{210} = -1/26873856 - 8 x/3,\  c_{213} = 1/21233664,\ c_{310} = 5/26873856,\\
&c_{311} = -5/15925248 - 3 y/8,\  c_{414} = 0,\ c_{412} = 413/1451188224 - 8 z/3,\\
&c_{416} = -7/120932352,\ c_{513} = -5/40310784,\  c_{512} = -3 w/2,\  d_{464} = 0.
\end{split}
\end{displaymath}

\end{document}